\documentclass[sigconf]{acmart}
\usepackage{subfigure}
\usepackage{graphicx}
\usepackage{booktabs}
\usepackage{multirow}
\usepackage{lipsum}
\usepackage{marvosym}
\usepackage{enumitem}

\AtBeginDocument{%
  \providecommand\BibTeX{{%
    \normalfont B\kern-0.5em{\scshape i\kern-0.25em b}\kern-0.8em\TeX}}}

\setcopyright{acmcopyright}
\copyrightyear{2022}
\acmYear{2022}

\setcopyright{acmcopyright}
\acmConference[CIKM '22] {Proceedings of the 31st ACM International Conference on Information and Knowledge Management}{October 17--21, 2022}{Atlanta, GA, USA.}
\acmBooktitle{Proceedings of the 31st ACM International Conference on Information and Knowledge Management (CIKM '22), October 17--21, 2022, Atlanta, GA, USA}
\acmPrice{15.00}
\acmISBN{978-1-4503-9236-5/22/10}
\acmDOI{10.1145/3511808.3557479}
\begin{document}

\title{Towards Understanding the Overfitting Phenomenon of Deep Click-Through Rate Prediction Models}


\author{Zhao-Yu Zhang}\authornote{Zhao-Yu Zhang, Xiang-Rong Sheng, and Yujing Zhang contributed equally to this research. This work was done when Zhao-Yu Zhang was a research intern at Alibaba Group, and Shuguang Han is the corresponding author.}
\email{zhangzhaoyu@lamda.nju.edu.cn}
\affiliation{%
  \institution{Nanjing University}
    \city{Nanjing}
    \country{China}
}

\author{Xiang-Rong Sheng}\authornotemark[1]
\email{xiangrong.sxr@alibaba-inc.com}
\author{Yujing Zhang}\authornotemark[1]
\email{jinghan.zyj@alibaba-inc.com}
\affiliation{%
  \institution{Alibaba Group}
    \city{Beijing}
    \country{China}
}

\author{Biye Jiang}
\email{biye.jby@alibaba-inc.com}
\author{Shuguang Han\textsuperscript{\Letter}}
\email{shuguang.sh@alibaba-inc.com}
\affiliation{%
  \institution{Alibaba Group}
    \city{Beijing}
    \country{China}
}

\author{Hongbo Deng}
\email{dhb167148@alibaba-inc.com}
\affiliation{%
  \institution{Alibaba Group}
    \city{Beijing}
    \country{China}
}

\author{Bo Zheng}
\email{bozheng@alibaba-inc.com}
\affiliation{%
  \institution{Alibaba Group}
    \city{Beijing}
    \country{China}
}

\renewcommand{\shortauthors}{Zhang, Sheng, and Zhang, et al.}

\begin{abstract}
Deep learning techniques have been applied widely in industrial recommendation systems. However, far less attention has been paid to the overfitting problem of models in recommendation systems, which, on the contrary, is recognized as a critical issue for deep neural networks. In the context of Click-Through Rate (CTR) prediction, we observe an interesting one-epoch overfitting problem: the model performance exhibits a dramatic degradation at the beginning of the second epoch. Such a phenomenon has been witnessed widely in real-world applications of CTR models. Thereby, the best performance is usually achieved by training with only one epoch. To understand the underlying factors behind the one-epoch phenomenon, we conduct extensive experiments on the production data set collected from the display advertising system of Alibaba. The results show that the model structure, the optimization algorithm with a fast convergence rate, and the feature sparsity are closely related to the one-epoch phenomenon. We also provide a likely hypothesis for explaining such a phenomenon and conduct a set of proof-of-concept experiments. We hope this work can shed light on future research on training more epochs for better performance.
\end{abstract}

\ccsdesc[500]{Information systems~Recommender systems}

\keywords{Deep Learning, Recommender Systems, Click-Through Rate Prediction, Overfitting, Sparse Feature}

\maketitle

\section{Introduction}
Deep learning techniques have driven advances in many application domains, ranging from computer vision~\cite{RussakovskyDSKS15Imagenet,HeZRS16ResNet}, natural language processing~\cite{DevlinCLT19Bert,VaswaniSPUJGKP2017Transformer} to recommender systems~\cite{CovingtonAS2016YouTubeDNN,cheng2016wide}. Along with its industrial prevalence, some research studies have investigated the overfitting problem of Deep Neural Networks (DNN)~\cite{Zhou2021WhyOverParamNotOverfit,salman2019overfitting,belkin2019reconciling}. These studies mainly experiment with image data sets, focusing on the connection between overfitting and model architectures~\cite{ZhangBHRV2021RethinkingGeneralization,salman2019overfitting}. However, less attention has been paid to the overfitting phenomenon of deep neural models for recommender systems. 

In this work, we study the overfitting problem of the deep click-through rate (CTR) prediction model~\cite{CovingtonAS2016YouTubeDNN,cheng2016wide,zhou2018din,zhou2019dien}. Despite the focus on CTR prediction, the analysis of this research can be easily generalized to other prediction tasks, such as deep conversion rate (CVR) prediction~\cite{Chapelle2014DFM,MaZHWHZG2018ESMM,GuSFZZ2021Defer}. For industrial applications, the CTR prediction task is commonly formulated as a supervised learning problem: a CTR prediction model fits the historical user-item click interactions in the training stage and then is evaluated on new user behaviors for testing.

\begin{figure}[t]
\includegraphics[width=.95\columnwidth]{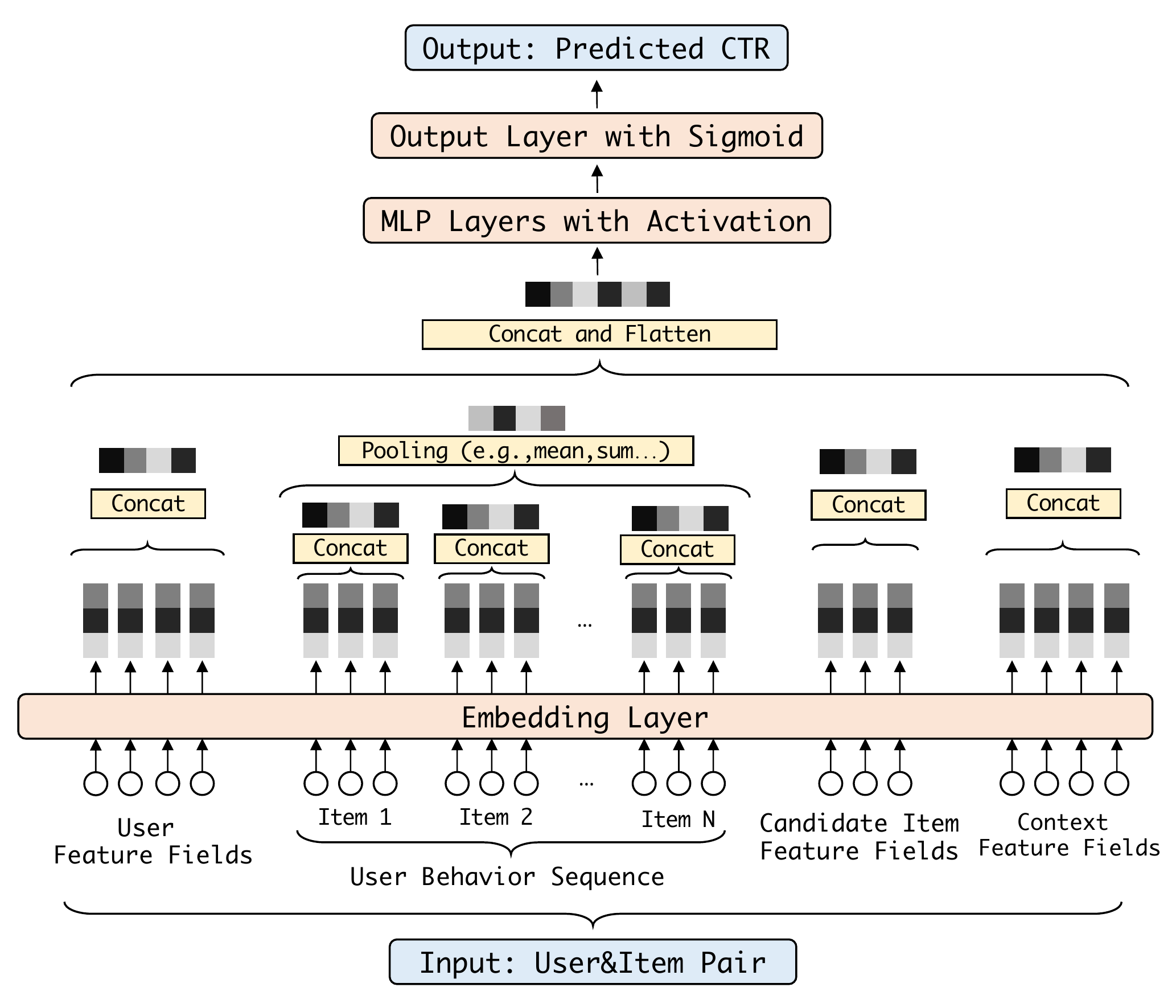}
\centering
\caption{An illustration of Embedding and MLP structure for the deep CTR prediction model.}
\label{fig:embedding_and_MLP}
\end{figure}

There are two main characteristics in CTR prediction, the \textbf{data} and the \textbf{model architecture}. Firstly, data in the recommender systems are high-dimensional and sparse. Deep models are trained on large-scale data sets with even billions of features but the vast majority of features have very low occurrences~\cite{Jiang2019XDL,zhang2022picasso, xie2020kraken}. Secondly, considering the data characteristic of recommender systems, CTR prediction models generally follow an Embedding and MLP architecture~\cite{zhou2018din}, unlike the common deep architecture like CNNs in computer vision. Figure~\ref{fig:embedding_and_MLP} illustrates the Embedding and MLP structure. The raw inputs, usually sparse category features represented by IDs, are first mapped into low-dimensional vectors by the embedding layer and then transformed into a fixed-length vector by pooling (e.g., mean
pooling) for each feature field (e.g., sequence of history clicked item IDs), and finally concatenated together as input of the following MLP layers for final prediction. 

\begin{figure}[t]
\includegraphics[width=.53\columnwidth]{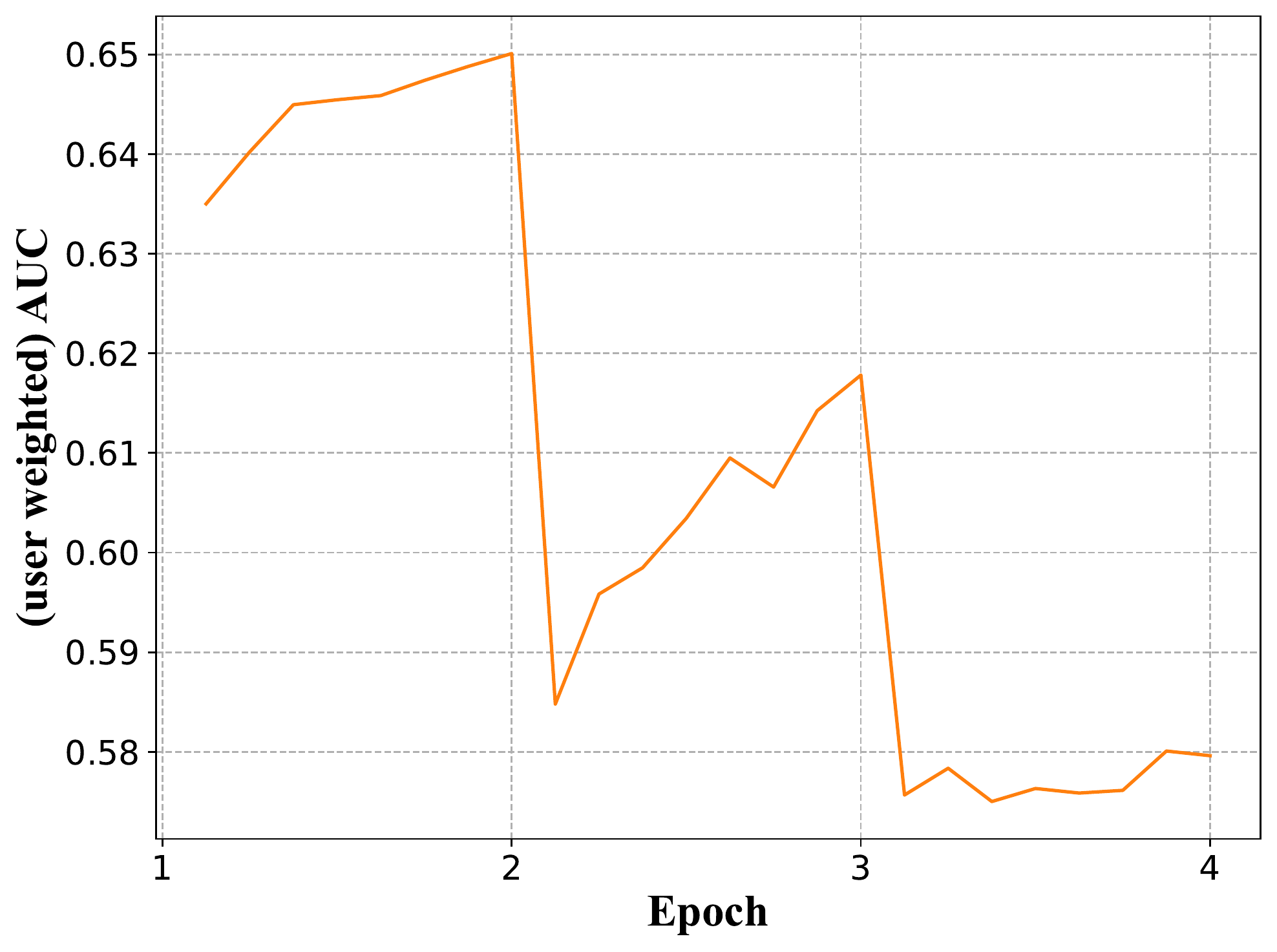}
\centering
\caption{An illustration of testing user-weighted AUC for our industrial CTR prediction model when training for multiple epochs. The model performance falls rapidly at the beginning of the second epoch.}
\label{fig:overfitting_online}

\end{figure}

By conducting extensive experiments on industrial recommender systems, we observe that the overfitting phenomenon of the deep CTR prediction model is peculiar. The model performance increases gradually within the first epoch while falls dramatically at the beginning of the second epoch (see Figure~\ref{fig:overfitting_online}). This phenomenon will be referred to as the \textbf{one-epoch phenomenon} in the following. It differs from overfitting in tasks like computer vision, where the model is usually trained for hundreds of epochs, and the overfitting occurs gradually. 
The finding of the one-epoch phenomenon of CTR prediction models is also consistent with previous experiments in industrial applications~\cite{zhou2018din} and academic research~\cite{zhu2021benchmarkCTR}. We believe a profound investigation of this topic will provide insights into understanding deep learning in recommender systems and drive forward the development of industrial CTR prediction models.

To this end, we conduct extensive experiments on public and production data sets to find the potential factors influencing the one-epoch phenomenon and offer insights into understanding it. The results show that the model structure, optimization algorithm with a fast convergence rate, and feature sparsity are closely related to the one-epoch phenomenon. We also discover that the one-epoch phenomenon exists widely in deep CTR prediction models. Moreover, we find training the models for more epochs does not improve over models trained in one epoch with appropriate hyper-parameters. Finally, we give a hypothesis to explain the one-epoch phenomenon and conduct proof-of-concept experiments, hoping to provide insights for the follow-up work. 

The main contributions are summarized as follows:
\begin{itemize}[leftmargin=*] 
    \item We conduct extensive experiments on industrial production data sets. The results show that the deep CTR prediction models exhibit the one-epoch phenomenon. Concretely, the models abruptly overfit the training data at the beginning of the second epoch, causing a severe drop in model performance.
    
    \item  We find that the model structure, optimization algorithm with a fast convergence rate, and feature sparsity are closely related to the one-epoch phenomenon. Although we can train the model for multiple epochs by restricting these factors, the best model performance is usually obtained by training only one epoch. The result may explain why most online industrial deep CTR prediction models only train the data once.
    
    \item  We provide a hypothesis to explain the one-epoch phenomenon and design experiments for verification. 
    Denote the representation after the embedding layer of sample $\boldsymbol{x}$ as $\text{EMB}(\boldsymbol{x})$. 
    The key points are that the joint probability distribution $\mathcal{D}\left(\text{EMB}(\boldsymbol{x}), y\right)$ is different between untrained and trained samples, and MLP layers quickly adapt to $\mathcal{D}\left(\text{EMB}(\boldsymbol{x}), y\right)$ of trained samples at the second epoch,   leading to the one-epoch phenomenon.
    
\end{itemize}



\section{Related Work}
\subsection{CTR Prediction}
\label{sec:ctr_prediction}
A fundamental goal in recommender systems is to recommend proper items to users, and an accurate CTR prediction model is critical for achieving this objective. In the deep learning era, many industrial CTR prediction models have made the transition from traditional shallow models~\cite{friedman2001greedy,rendle2010factorization,KorenBV09MF4RS} to deep models~\cite{cheng2016wide,guo2017deepfm,QuFZTNGYH19PNN,zhou2018din,zhou2019dien,ShengZZDDLYLZDZ2021STAR}. As shown in Figure~\ref{fig:embedding_and_MLP}, a deep CTR prediction model generally follows an Embedding and MLP architecture. The embedding module first transforms each discrete ID from raw input into low dimensional vector. The embeddings of each feature field are then aggregated by various means (e.g., mean pooling) to obtain a fixed-length vector. The embedding vectors of different feature fields are concatenated as input into the MLP module for final prediction. Under this architecture, there are many studies on improving some of the components. For example, studies on user interest modeling~\cite{zhou2018din,zhou2019dien} focus on effective ways of aggregating user behaviors. Feature interaction~\cite{QuFZTNGYH19PNN} mainly focuses on the interaction of embedding vectors of different feature fields to generate high-order features and concatenate them into the input vectors of the MLP layers. The Embedding and MLP architecture has achieved state-of-the-art performance and has been deployed widely in industrial recommender systems.

\subsection{Overfitting Phenomenon of DNN}
Deep learning has achieved state-of-the-art performances in many application domains. Along with its empirical success, many researchers attempt to understand the overfitting phenomenon of DNN. Zhou~\cite{Zhou2021WhyOverParamNotOverfit} regards the DNN as the combination of the feature space transformation (FST) part and classifier construction (CC) part. Over-parameterization leads to overfitting in CC but not in FST. Salman and Liu~\cite{salman2019overfitting} show that the overfitting of DNN is due to continuous gradient updating and scale sensitiveness of cross-entropy loss. In addition, there are some studies on the generalization ability of DNN, which is closely related to overfitting. Zhang et al.~\cite{ZhangBHRV2021RethinkingGeneralization} observe that conventional generalization bounds are inadequate for over-parameterized DNN. Bartlett et al.~\cite{BartlettFT2017SpectralMarginBound} present a margin-based generalization bound for neural networks that scale with their margin-normalized spectral complexity. In conclusion, despite many exploratory studies, there is currently no widely-accepted explanation or widely-used theoretical tool, and understanding the overfitting of DNN remains an open problem.

Unlike models in other tasks, deep models in recommender systems usually follow the Embedding and MLP structure, facing high-dimensional sparse feature~\cite{Jiang2019XDL,zhang2022picasso}. The data involve billions of sparse features, and only part of the parameters are used in each forward pass. It brings new challenges to applying deep learning algorithms and analyzing their overfitting phenomenon. To the best of our knowledge, far less attention is paid to the overfitting phenomenon of deep models in recommender systems.

\section{The One-Epoch Phenomenon}
To better illustrate the overfitting problem of deep CTR prediction models, we show the testing curve of the online CTR prediction model of Alibaba display advertising platform. The model utilizes hundreds of feature fields including raw IDs (e.g., ID or categorical attribute of an item), interaction features, and statistical features (e.g., historical stats of a user or item). As shown in Figure~\ref{fig:overfitting_online}, the testing user-weighted AUC (Area Under the ROC Curve) drops dramatically at the beginning of the second epoch (and the beginning of other epochs as well). We name it the one-epoch phenomenon.

\begin{figure}[t]
\includegraphics[width=.53\columnwidth]{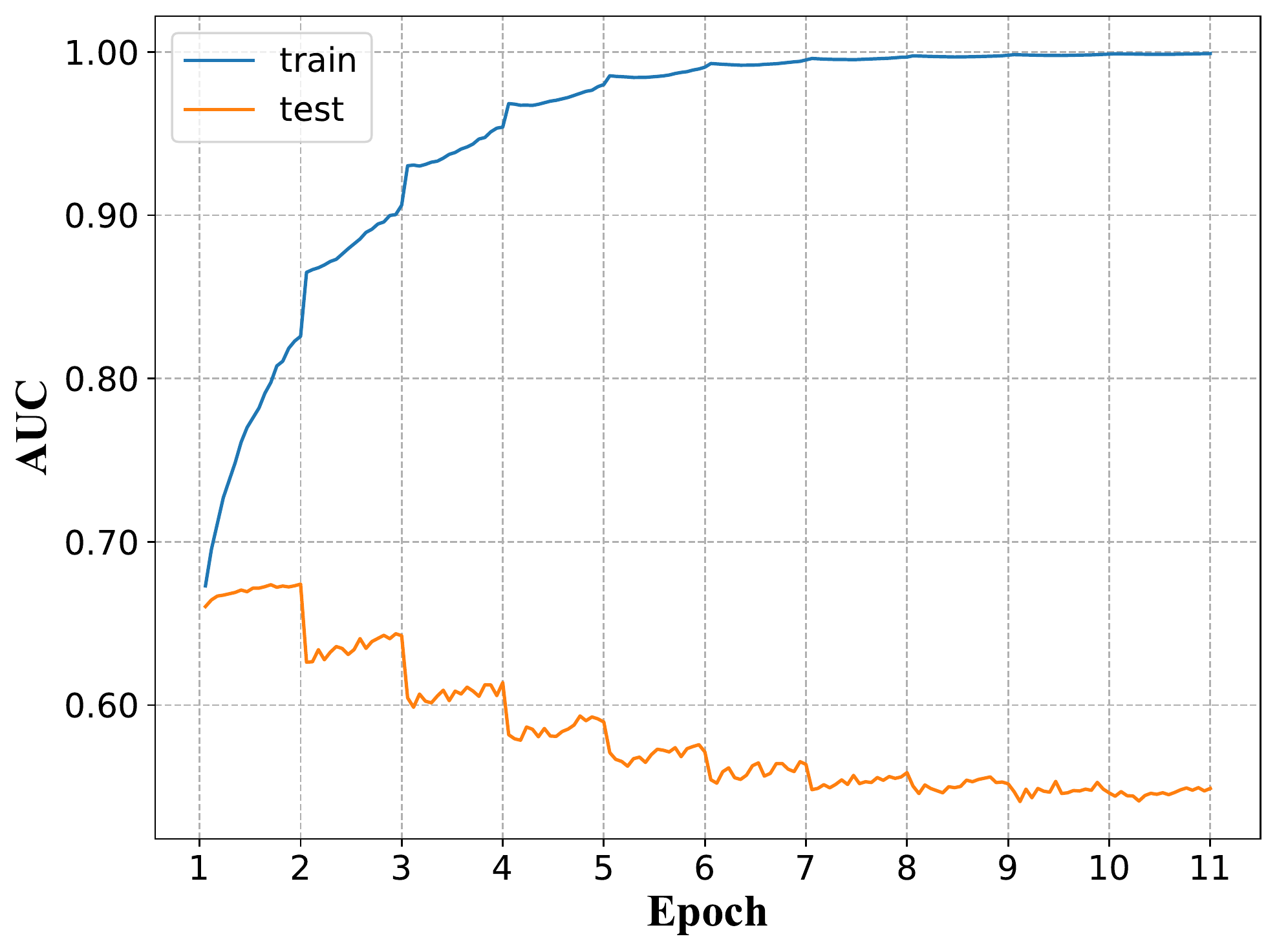}
\centering
\caption{An illustration of the training and testing AUC for the production data set over different training epochs. Here, we can clearly observe the one-epoch phenomenon.}
\label{fig:overfitting_production}
\end{figure}

Since the industrial CTR prediction model is specifically tailored and optimized for our online applications, we further conduct a set of experiments with other commonly-adopted model architectures and a simplified production data set. We firstly examine the simplest DNN structure -- a three-layer DNN with hidden units (200, 80, 1), and seven simple feature fields (as shown in Table~\ref{tab:production_feature_field_entropy}) are adopted for experiments. Figure~\ref{fig:overfitting_production} plots the model performance for ten epochs. It is clear that the one-epoch phenomenon remains, even with the simplest DNN model structure and input features.

\begin{figure}[t]
\includegraphics[width=.53\columnwidth]{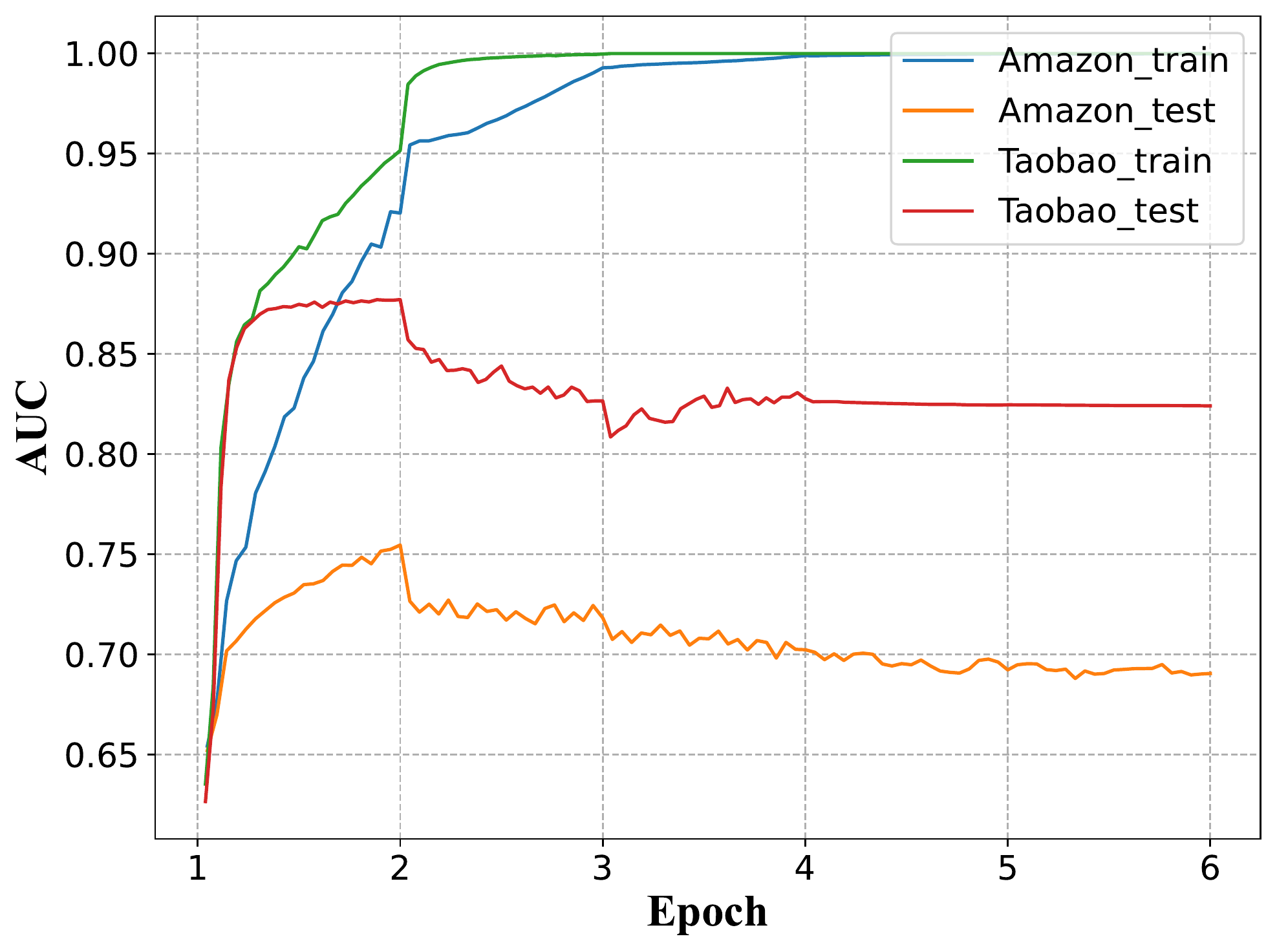}
\centering
\caption{An illustration of the training and testing AUC for Amazon book and Taobao data sets over different training epochs. The one-epoch phenomenon occurs in both data sets.}
\label{fig:overfitting_others}
\end{figure}

In addition to the production data set, we also examine the overfitting phenomenon on two widely-used public data sets (Amazon book\footnote{http://jmcauley.ucsd.edu/data/amazon/} and Taobao\footnote{https://tianchi.aliyun.com/dataset/dataDetail?dataId=649}). Figure~\ref{fig:overfitting_others} offers an illustration of model performance over the number of training epochs, in which the one-epoch phenomenon still exists. This clearly demonstrates that the one-epoch phenomenon is a common problem, and it is not only restricted to one specific model structure or data set.

Similar observations are also made in previous studies~\cite{zhu2021benchmarkCTR,zhou2018din} on CTR prediction. Zhu et al.~\cite{zhu2021benchmarkCTR} discover that the best performance of many deep CTR prediction models on the Avazu data set is obtained by training only one epoch. Zhou et al.~\cite{zhou2018din} observe that the CTR prediction model performance abruptly decreases at the beginning of the second epoch. Despite the pervasiveness of the one-epoch phenomenon for deep CTR prediction models, to the best of our knowledge, no previous studies have been devoted to understanding such a phenomenon.

\section{The Anatomy of One-Epoch Phenomenon}\label{sec:analysis}
We conduct extensive experiments on the production data set to understand the likely factors that cause the one-epoch phenomenon. The experiments are divided into two parts: model-related factors and feature-related factors. As for the experiments, all of them are conducted on our production data set. Unless otherwise specified, the hyper-parameters of the model are the default settings. Details about the experiment settings are provided in Section~\ref{sec:experiment}. And experimental results on public data sets are provided with the codes for reproducibility.

\begin{figure*}[ht]

\centering
\subfigure[Embedding dimension]{
\captionsetup{width=0.7\columnwidth}

\includegraphics[width=.42\columnwidth]{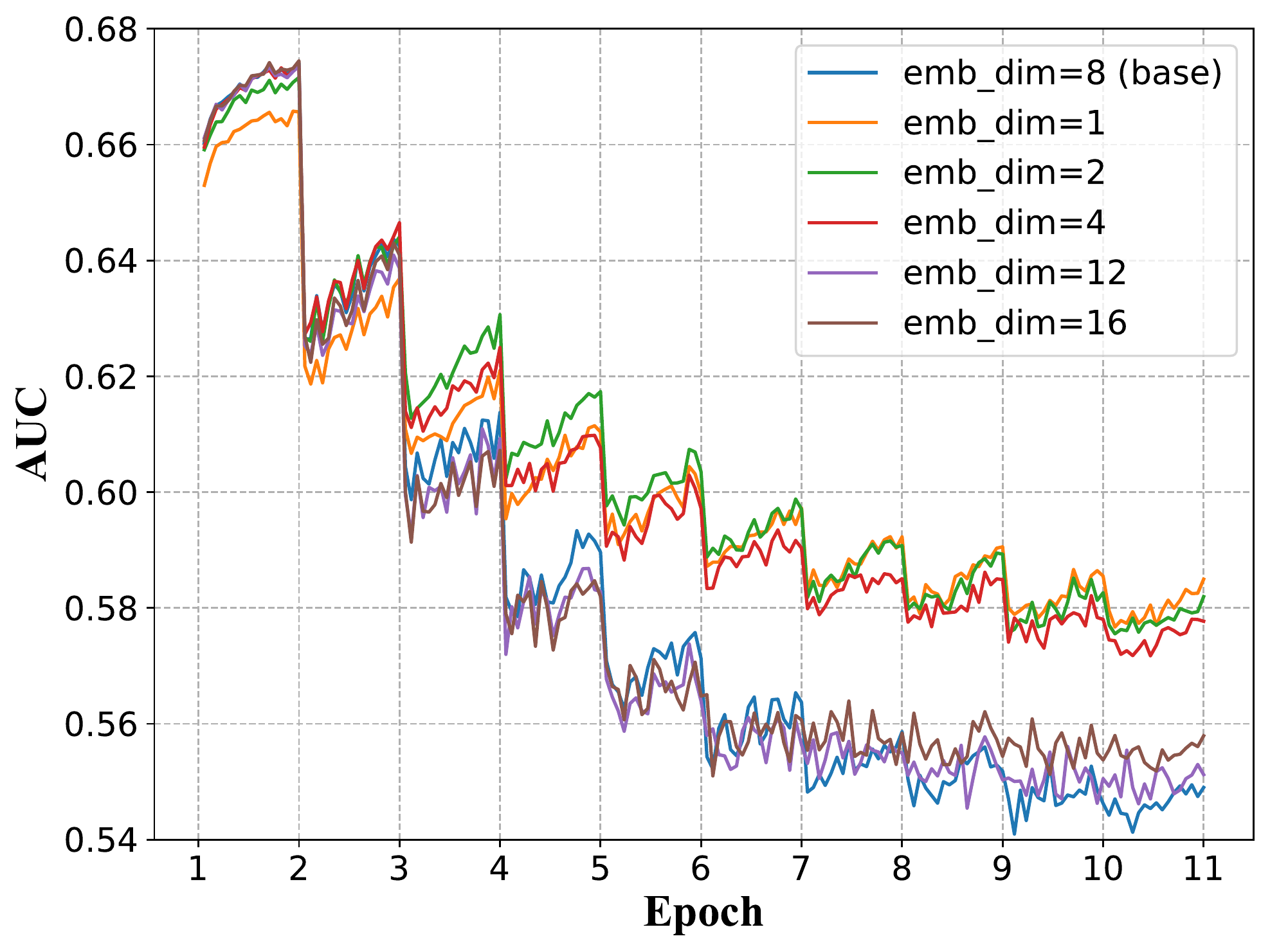}
\label{fig:model_structure_emb_dim_production}
}
\quad\quad\quad
\subfigure[Number (``$no$.") of MLP neurons.]{
\captionsetup{width=2\columnwidth}
\label{fig:model_structure_neuron_production}

\includegraphics[width=.42\columnwidth]{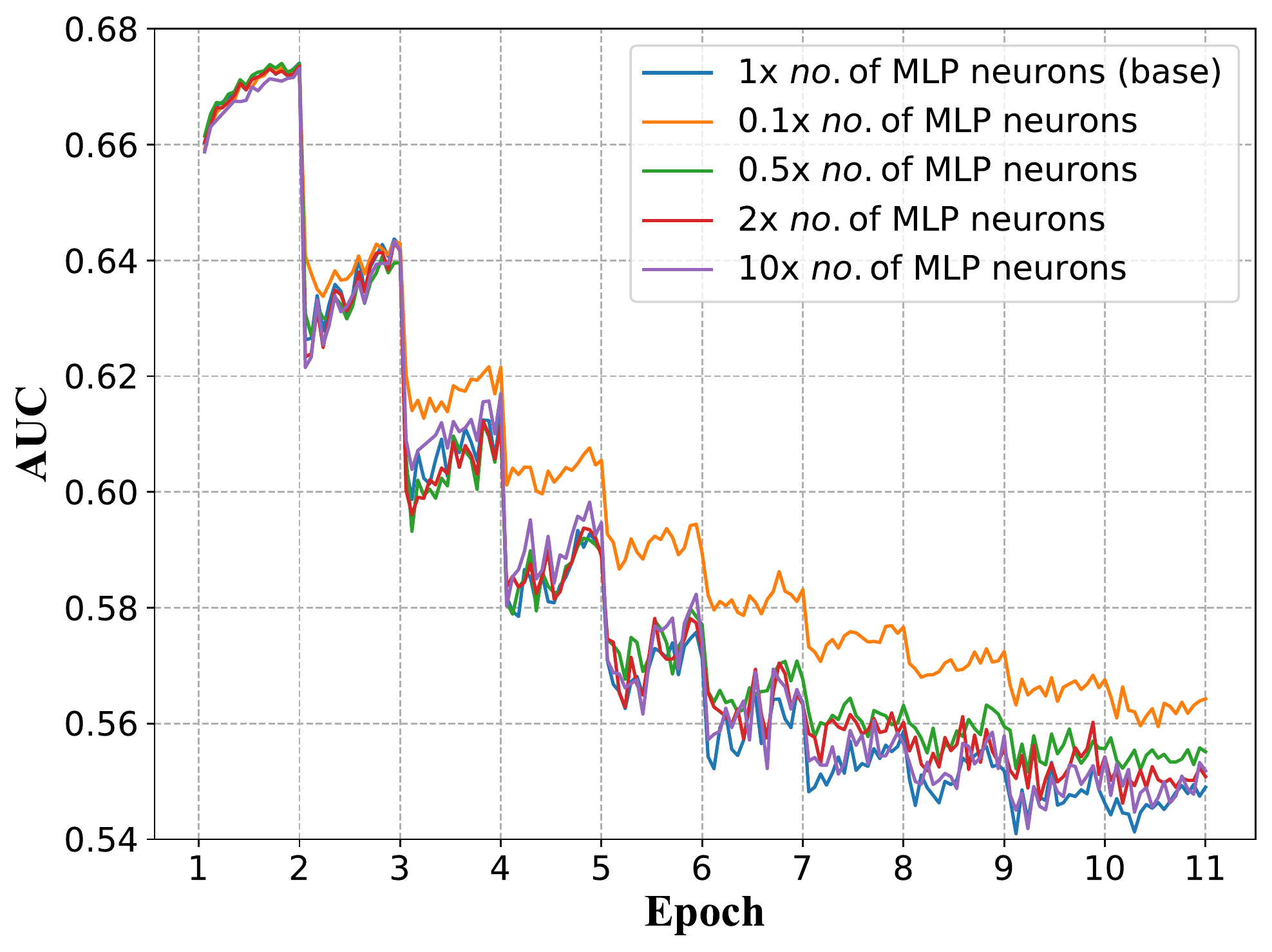}
}
\quad\quad\quad
\subfigure[Number of MLP layers.]{
\label{fig:model_structure_nlayers_production}
\includegraphics[width=.42\columnwidth]{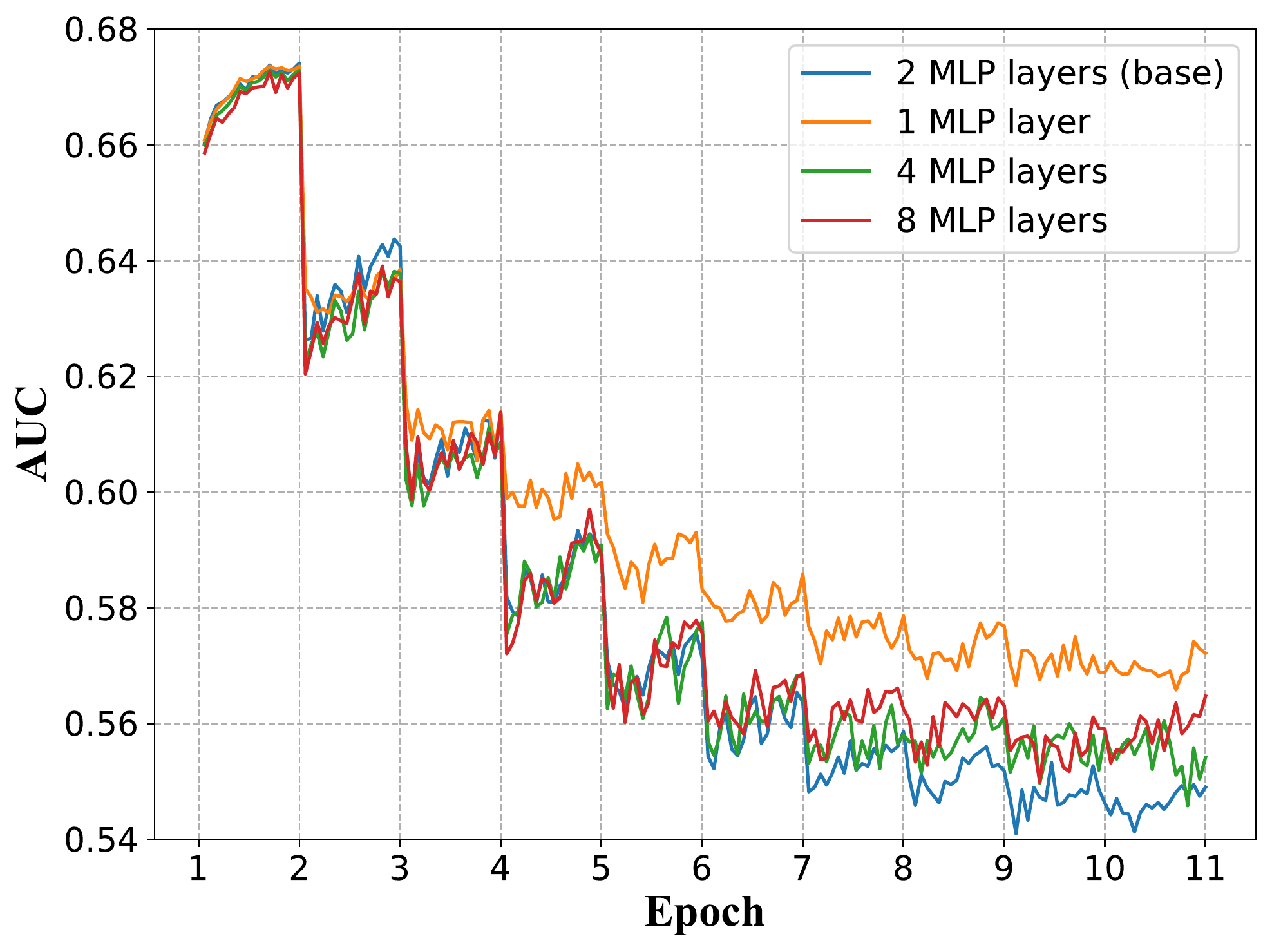}
}

\centering
\caption{The amount of model parameters (including the embedding dimension, the number of hidden units, and the number of MLP layers) has no clear effect on the one-epoch phenomenon. }
\label{fig:model_structure_parameters}

\end{figure*}

\subsection{Model-Related Factors}
\label{sec:basic_factors}

In this section, we study the impact of various model-related factors, ranging from the model architecture, number of model parameters, batch size, activation function, to the choice of training optimizer and learning rate, on the one-epoch phenomenon. In addition to that, we also experiment with multiple commonly-adopted techniques such as weight decay~\cite{loshchilov2017adamw} and dropout~\cite{srivastava2014dropout}, aiming to alleviate the overfitting problem.

\begin{figure}[t]
\includegraphics[width=.53\columnwidth]{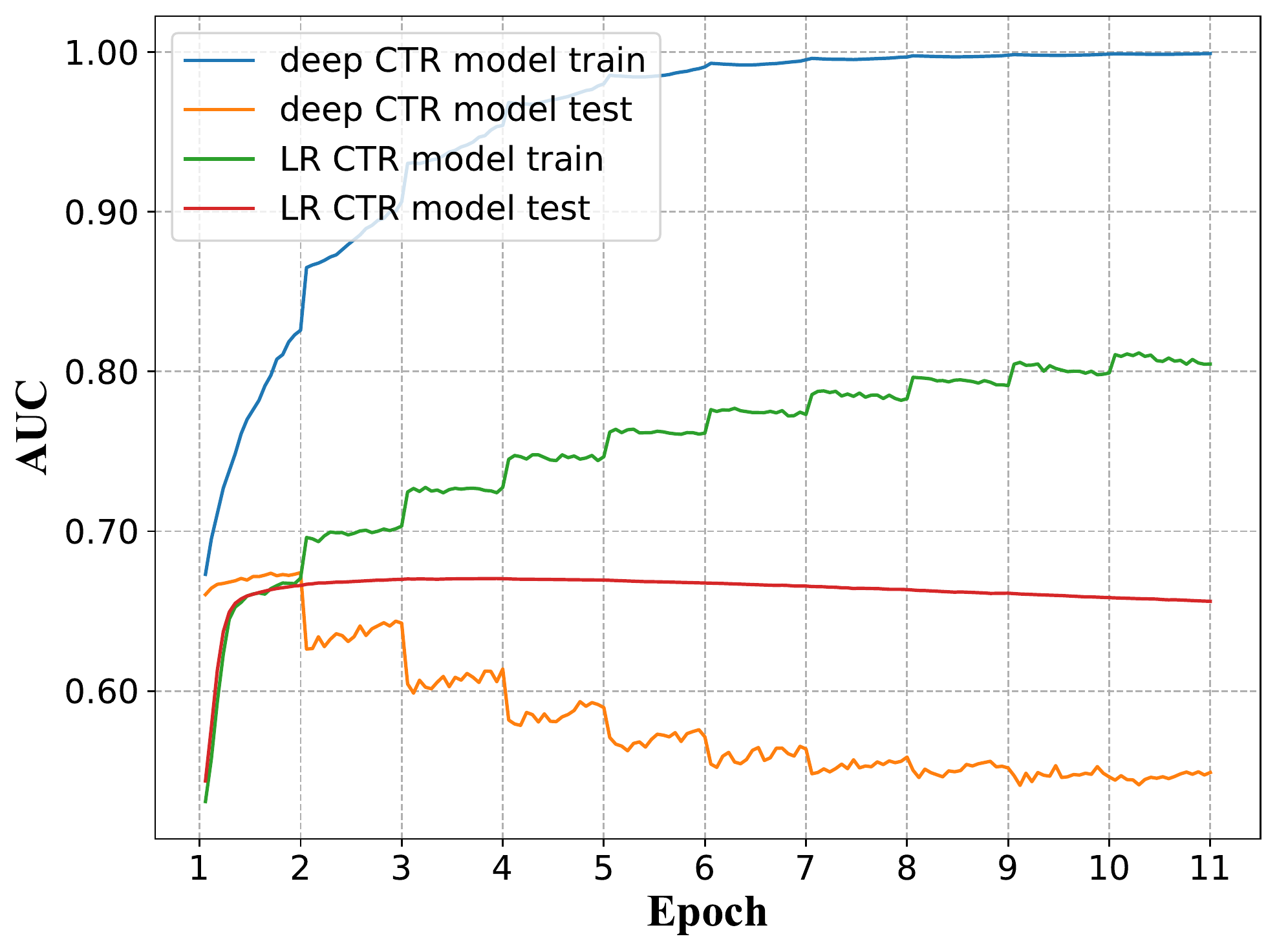}
\centering
\caption{A comparison of the convergence curves between the DNN model and the logistic regression (LR) model, in which the one-epoch phenomenon is only observed for DNN.}
\label{fig:model_structure_DNN_LR}
\end{figure}

\subsubsection{Model Structure}
For CTR prediction, Figure~\ref{fig:model_structure_DNN_LR} provides a comparison of the convergence curves between the DNN model and the LR model. In spite of slight overfitting, LR does not show the one-epoch phenomenon. It is worth noting that we have also experimented with a variety of parameter settings (e.g., learning rate, optimizer) for the LR model, and the model convergence curves present very similar patterns. On the contrary, the DNN-based models exhibit a clear one-epoch overfitting phenomenon. This illustrates that the one-epoch phenomenon is closely related to the adopted deep neural model architecture that consists of the utilization of embedding features and the corresponding MLP structure.

\subsubsection{The Amount of Model Parameters}
In addition to the default setups for embedding dimension (8 in the above experiments) and hidden units of the MLP layer, we conduct an extensive amount of experiments by varying model parameters. In this section, we analyze the effect of model parameters on the one-epoch phenomenon.

Figure~\ref{fig:model_structure_emb_dim_production} plots the model convergence curves over a variety of embedding dimension sizes, Figure~\ref{fig:model_structure_neuron_production} illustrates the effect of model convergence with different number of hidden units, and Figure~\ref{fig:model_structure_nlayers_production} compares the model performance for different number of MLP layers. It is clear that with the adopted deep CTR prediction model, different setups for model parameters do not mitigate such a one-epoch phenomenon. 

For a more special case where the embedding dimension size is set to 1, meaning that every feature is only represented by one scalar value, the total amount of parameters is roughly the same as the LR model. We observe that, even in this case, such a DNN-based model still suffers from the one-epoch phenomenon. Thereby, we believe that \textbf{the Embedding and MLP model structure rather than the amount of model parameters is a more related factor to the one-epoch phenomenon}.

\subsubsection{Activation Function}
\begin{figure}[t]
\subfigure[Batch size]{
\includegraphics[width=.42\columnwidth]{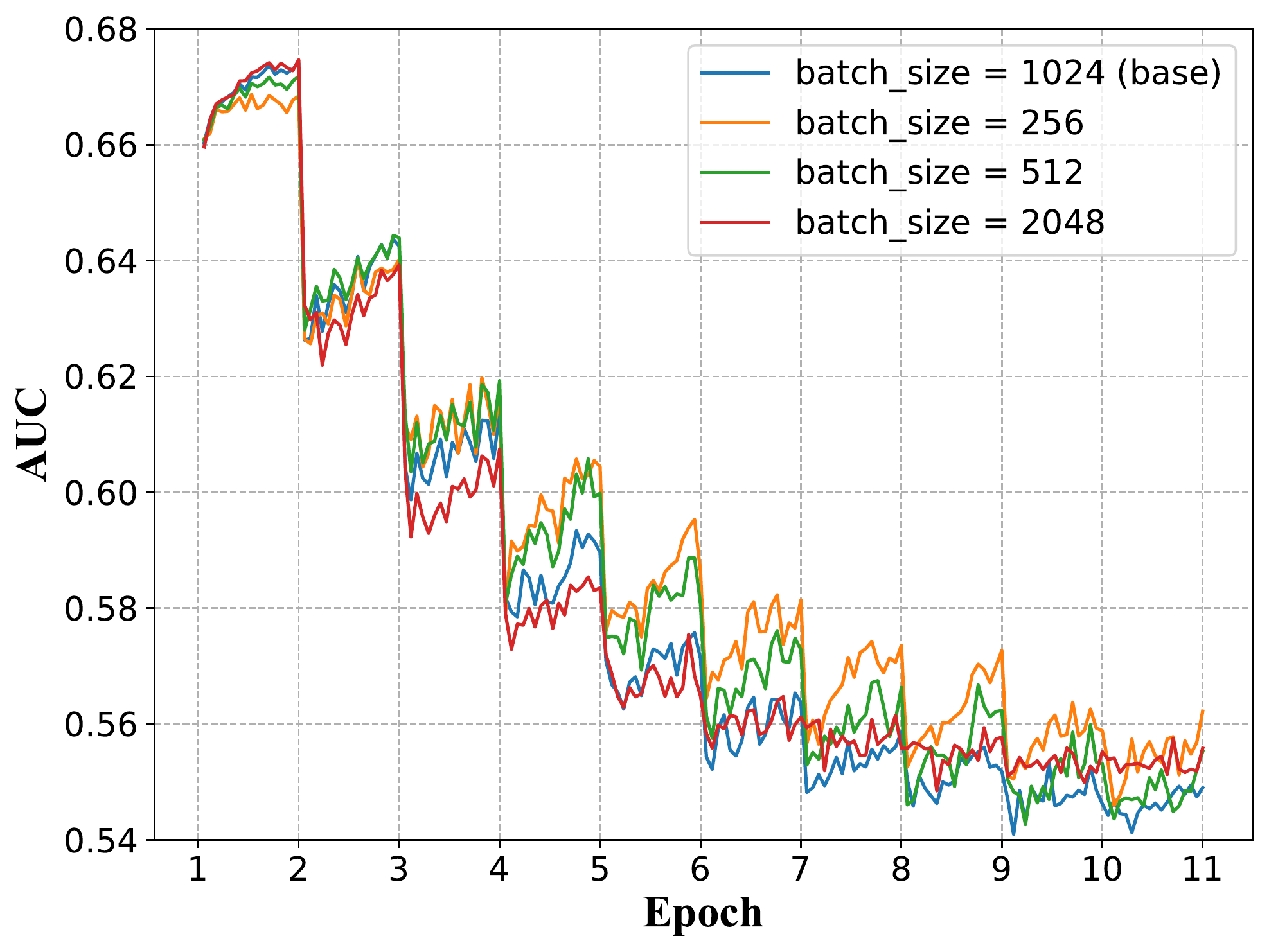}
\label{fig:model_structure_batch_size}
}
\subfigure[Activation function]{
\includegraphics[width=.42\columnwidth]{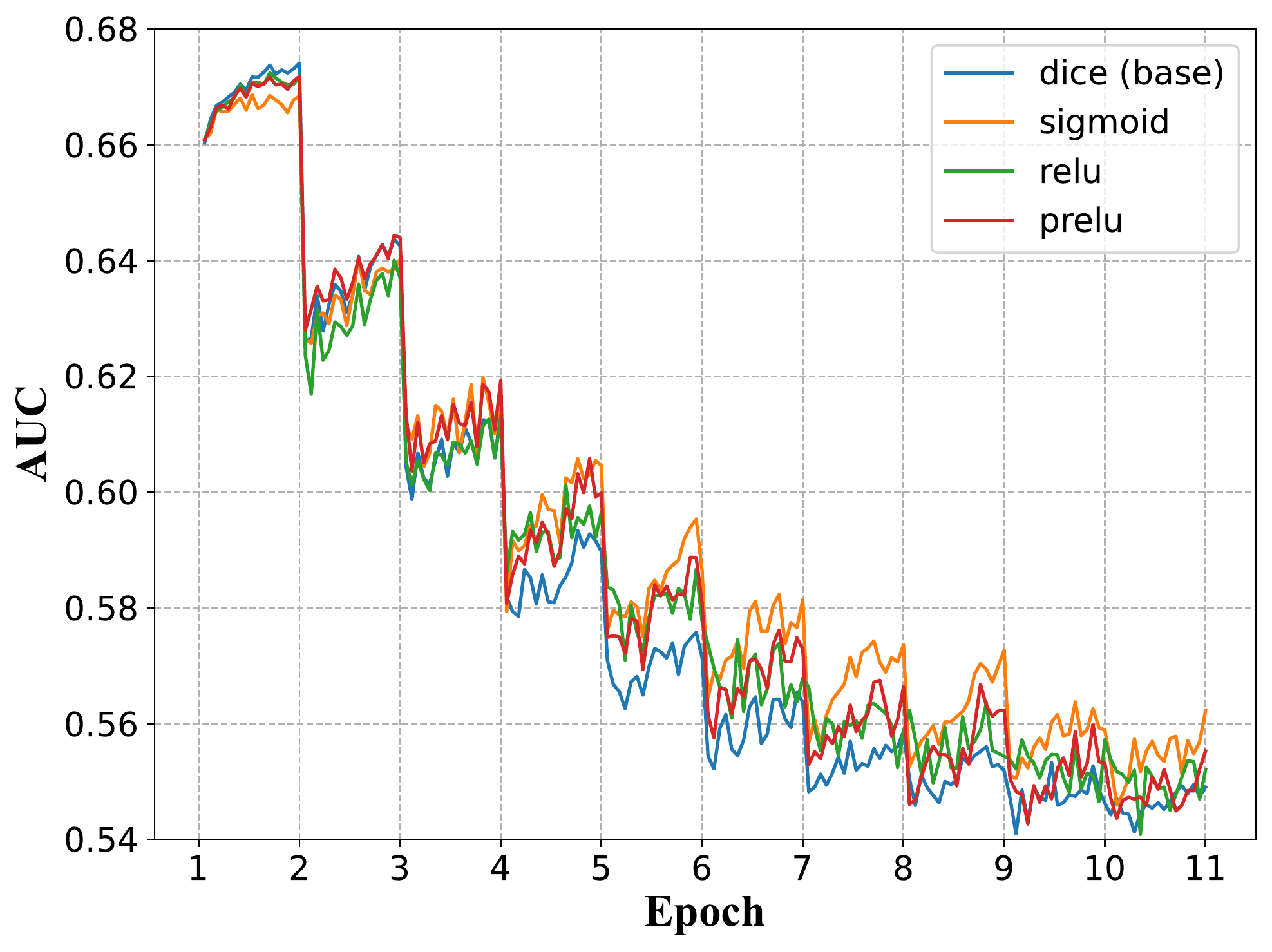}
\label{fig:model_structure_activation}
}
\centering
\caption{A comparison of testing AUC for models with different batch sizes and activation functions. Neither shows a significant effect on the one-epoch phenomenon.}
\label{fig:model_structure_batch_size_activation}

\end{figure}

We study the effect of different activation functions on the one-epoch phenomenon. In addition to the dice unit~\cite{zhou2018din}, we also employ sigmoid, relu~\cite{dahl2013relu} and prelu~\cite{he2015prelu}. Figure~\ref{fig:model_structure_activation} provides the model convergence curves for different activation functions, in which we find that the activation function almost has no influence on such a phenomenon.

\subsubsection{Batch Size}
We also analyze the effect of different batch sizes in Figure~\ref{fig:model_structure_batch_size}. Same to the activation function, changing batch sizes does not help alleviate the one-epoch problem.

\begin{figure*}[ht]
\centering
\subfigure[Adam]{
\includegraphics[width=.42\columnwidth]{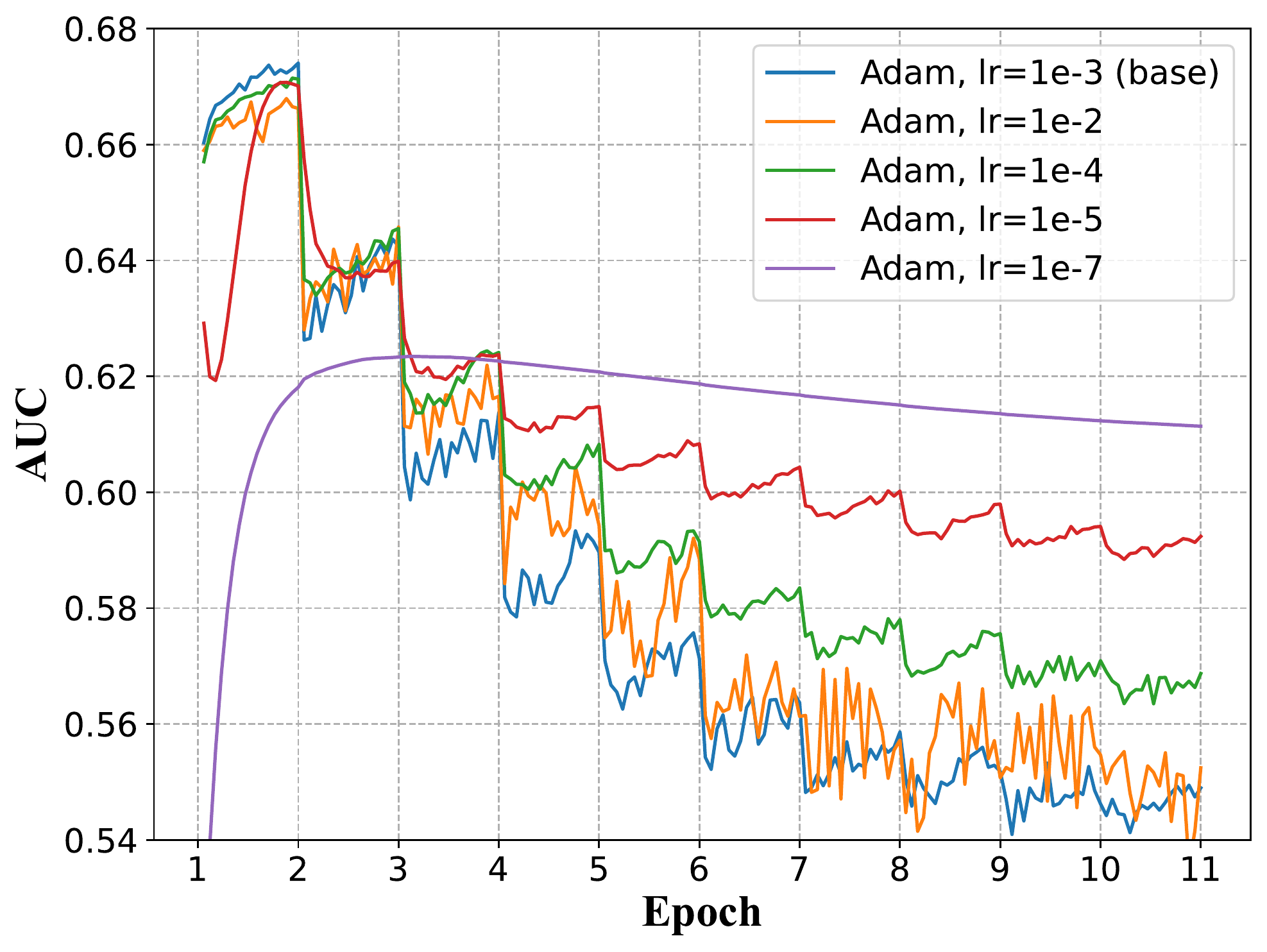}
\label{fig:model_optimizer_adam}
}
\quad\quad\quad
\subfigure[RMSprop]{
\includegraphics[width=.42\columnwidth]{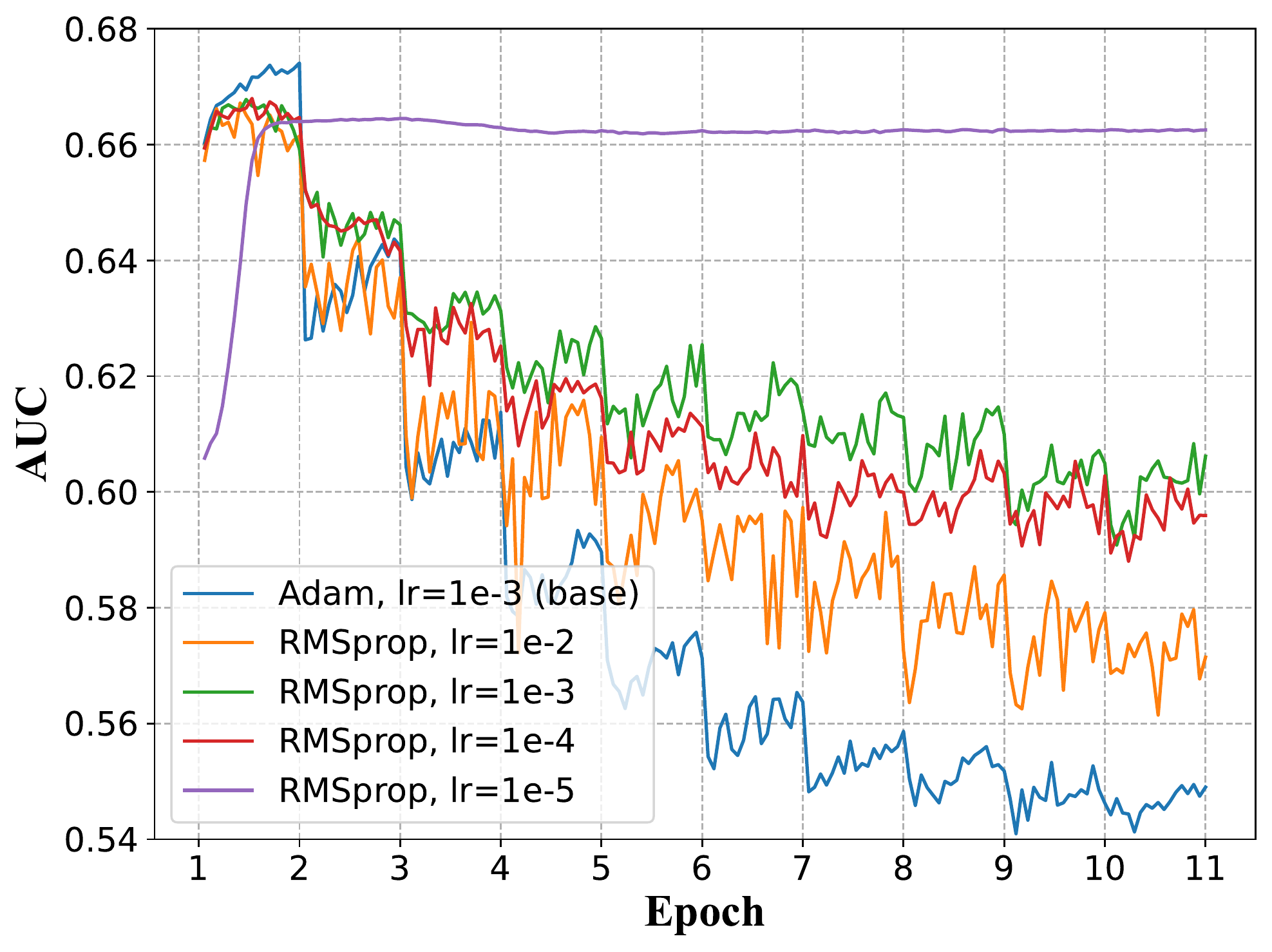}
}
\quad\quad
\subfigure[SGD]{
\includegraphics[width=.42\columnwidth]{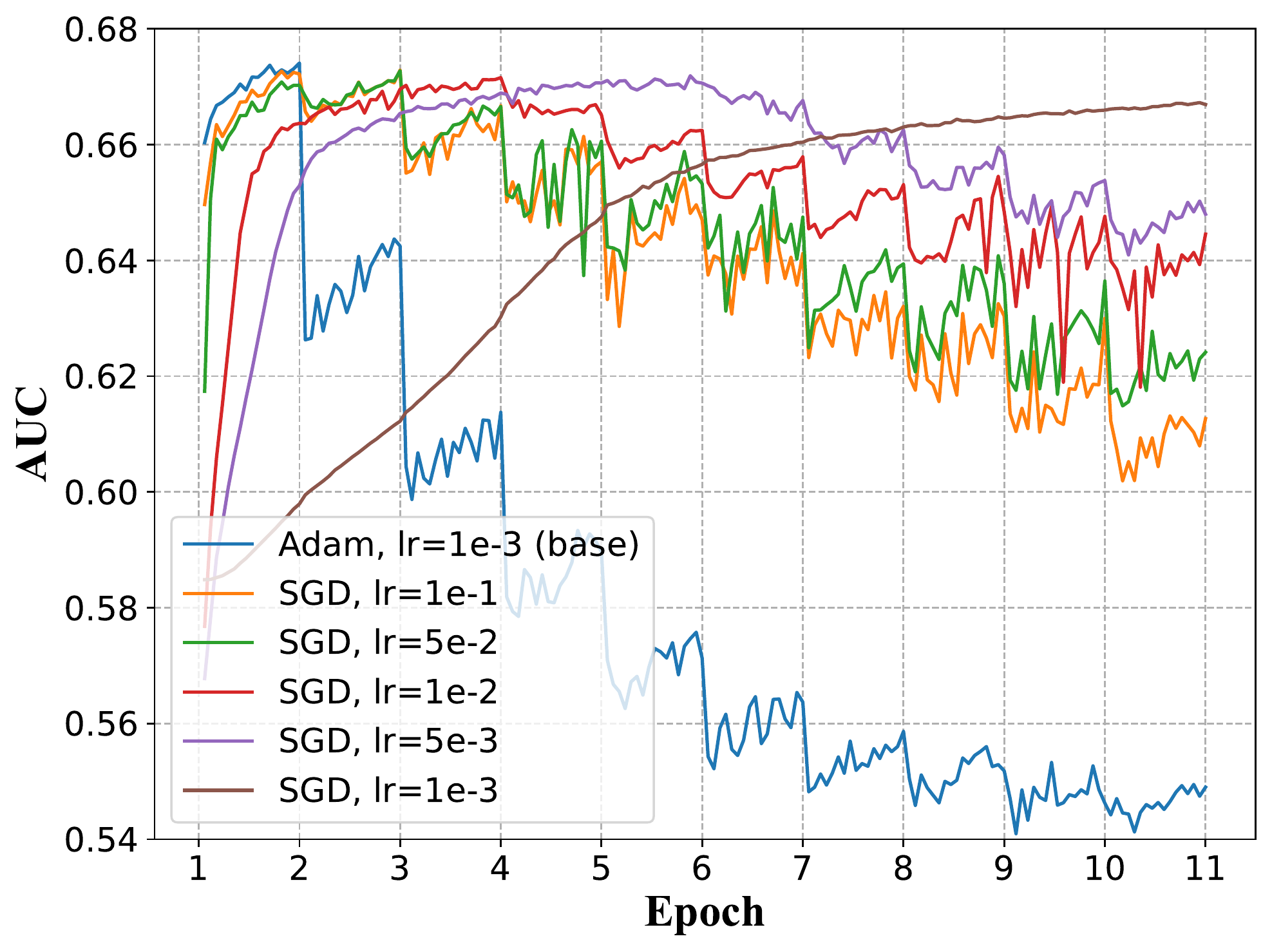}
}

\centering
\caption{A comparison of testing AUC for models with different optimizers. ``lr" is short for learning rate.}
\label{fig:model_optimizer}
\end{figure*}

\subsubsection{Optimization Algorithm} In this section, we focus on how optimization algorithms (including training optimizer and learning rate) affect the one-epoch phenomenon. In addition to the widely-adopted Adam optimizer~\cite{KingmaB2014Adam}, we further take into account RMSprop~\cite{hinton2012rmsprop} and Stochastic Gradient Descent (SGD). Figure~\ref{fig:model_optimizer} presents the model convergence curves of each optimizer over different training epochs. Compared to SGD, Adam and RMSprop show faster convergence rates in most of the cases, while they are more prone to the one-epoch phenomenon. We further observe that the learning rate may also be connected with the one-epoch problem. With an extremely small learning rate, such a phenomenon is less obvious, but it is at the expense of model performance. In summary, \textbf{an optimization algorithm that facilitates a faster model convergence could be at the risk of the one-epoch problem}.

\subsubsection{Weight Decay and Dropout}
\begin{figure}[t]
\subfigure[Weight decay. $\lambda$ is the weight decay factor defined in Equation~\ref{eqn:weight_decay}]{
\includegraphics[width=.42\columnwidth]{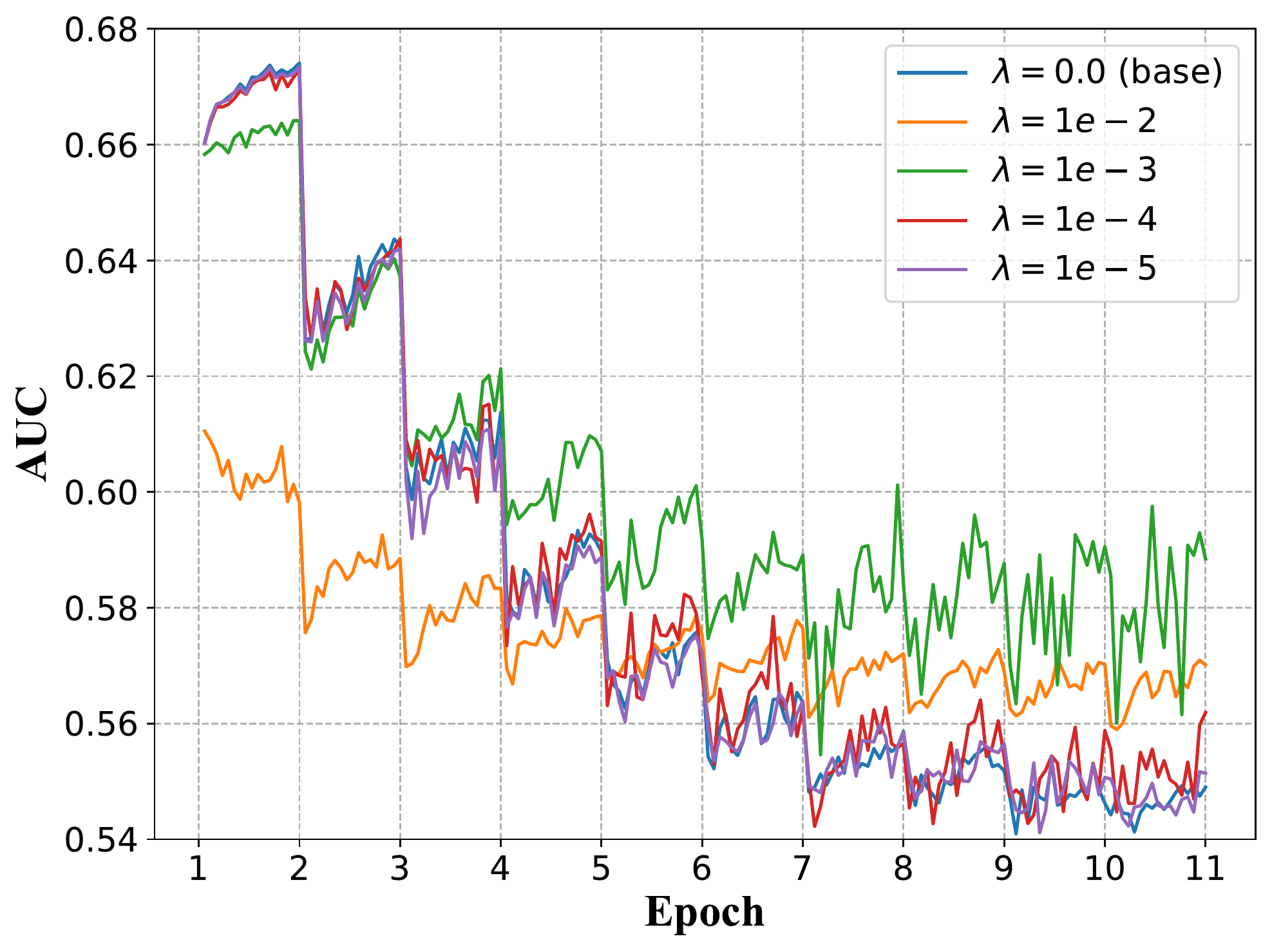}
\label{fig:model_structure_weight_decay}
}
\subfigure[Dropout]{
\includegraphics[width=.42\columnwidth]{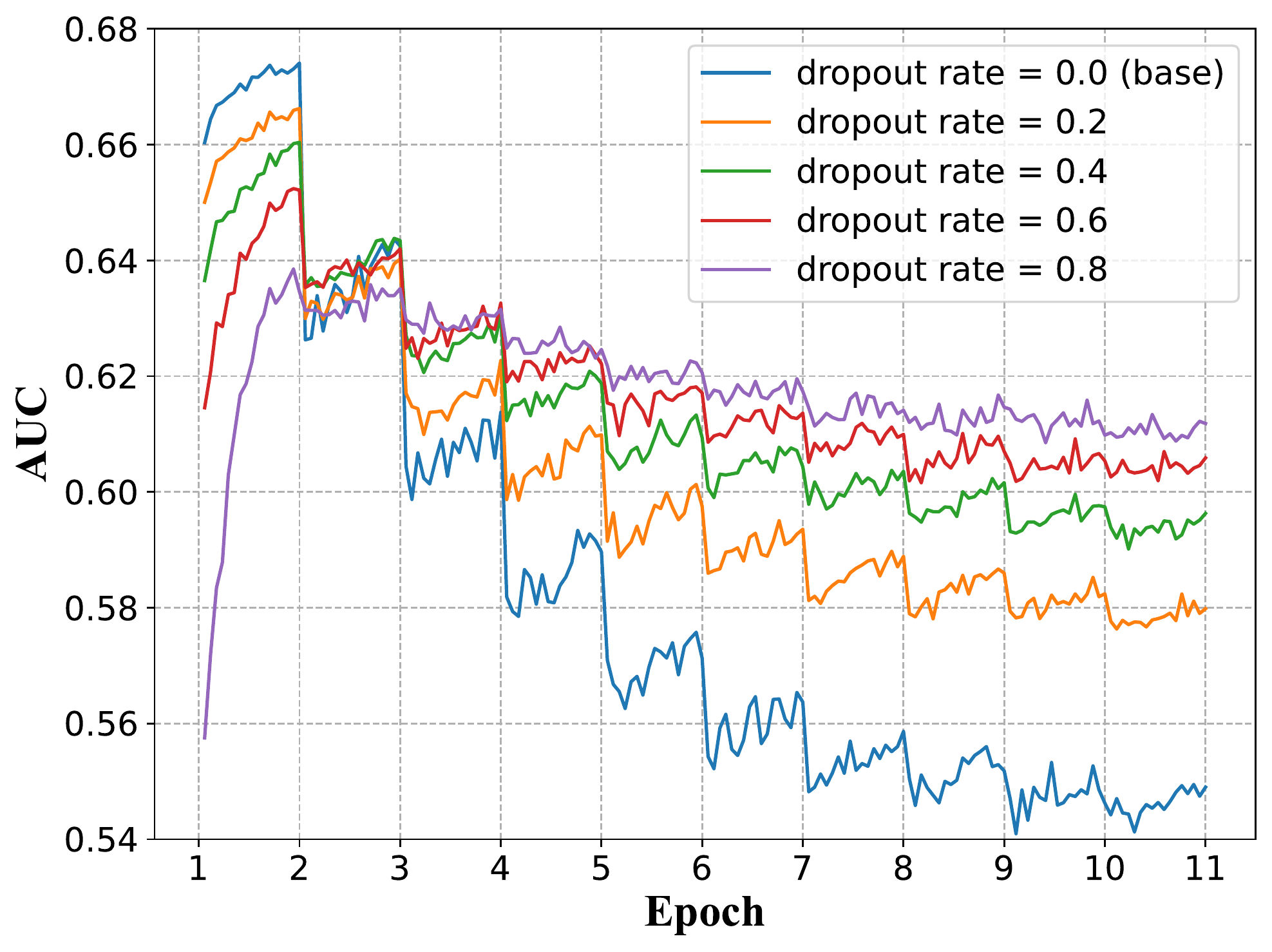}
\label{fig:model_structure_dropout}
}
\centering
\caption{A comparison of testing AUC for models with different weight decay and dropout settings. They do not show any sign of mitigating the one-epoch phenomenon.}
\label{fig:model_structure_weight_decay_dropout}
\end{figure}

Weight decay~\cite{loshchilov2017adamw} is a widely-adopted technique to restrict model complexity and thereby alleviate overfitting. At each iteration of the model training process, the weights of a neural network $\boldsymbol{\theta}$ are commonly updated with the computed gradient $\nabla f(\boldsymbol{\theta})$, whereas the weight decay technique further shrinks $\boldsymbol{\theta}$ with the below formula:
\begin{equation}
\label{eqn:weight_decay}
    \boldsymbol{\theta}_{t+1} = \boldsymbol{\theta}_{t} - \alpha \nabla f(\boldsymbol{\theta}_t) - \lambda \boldsymbol{\theta}_{t},
\end{equation}where $\alpha$ stands for the learning rate and $\lambda$ denotes the weight decay factor. Figure~\ref{fig:model_structure_weight_decay} plots the convergence curves over different choices of weight decay factors. We find that the weight decay algorithm neither improves the model performance nor helps alleviate the one-epoch problem.

Dropout~\cite{srivastava2014dropout} is another common technique to relieve the overfitting problem. The key idea is to randomly drop units (along with their connections) from the neural network during training, which prevents the units from co-adapting too much. In our experiment, we adopt one dropout layer before each fully connected layer, so there are 3 dropout layers in total. Figure~\ref{fig:model_structure_dropout} presents the model performance after applying the dropout layers, with each curve indicating one dropout rate. Again, we observe that dropout does not help solve the one-epoch problem.

\subsection{Feature-Related Factors}
\label{sec:data_sparsity}
The feature set used in industrial CTR prediction models can be roughly categorized into four types: user feature fields (e.g., age and gender), user behavior sequences (e.g., a sequence of clicked items), candidate item feature fields (e.g., item ID and category ID), and contextual feature fields. Table~\ref{tab:production_feature_field_entropy} gives an example of the production data set. Note that all features are preprocessed into discrete features (continuous features are discretized by buckets) and each feature value is represented by an ID, which is a common practice in industrial scenarios.

\begin{figure}[t]
\includegraphics[width=.53\columnwidth]{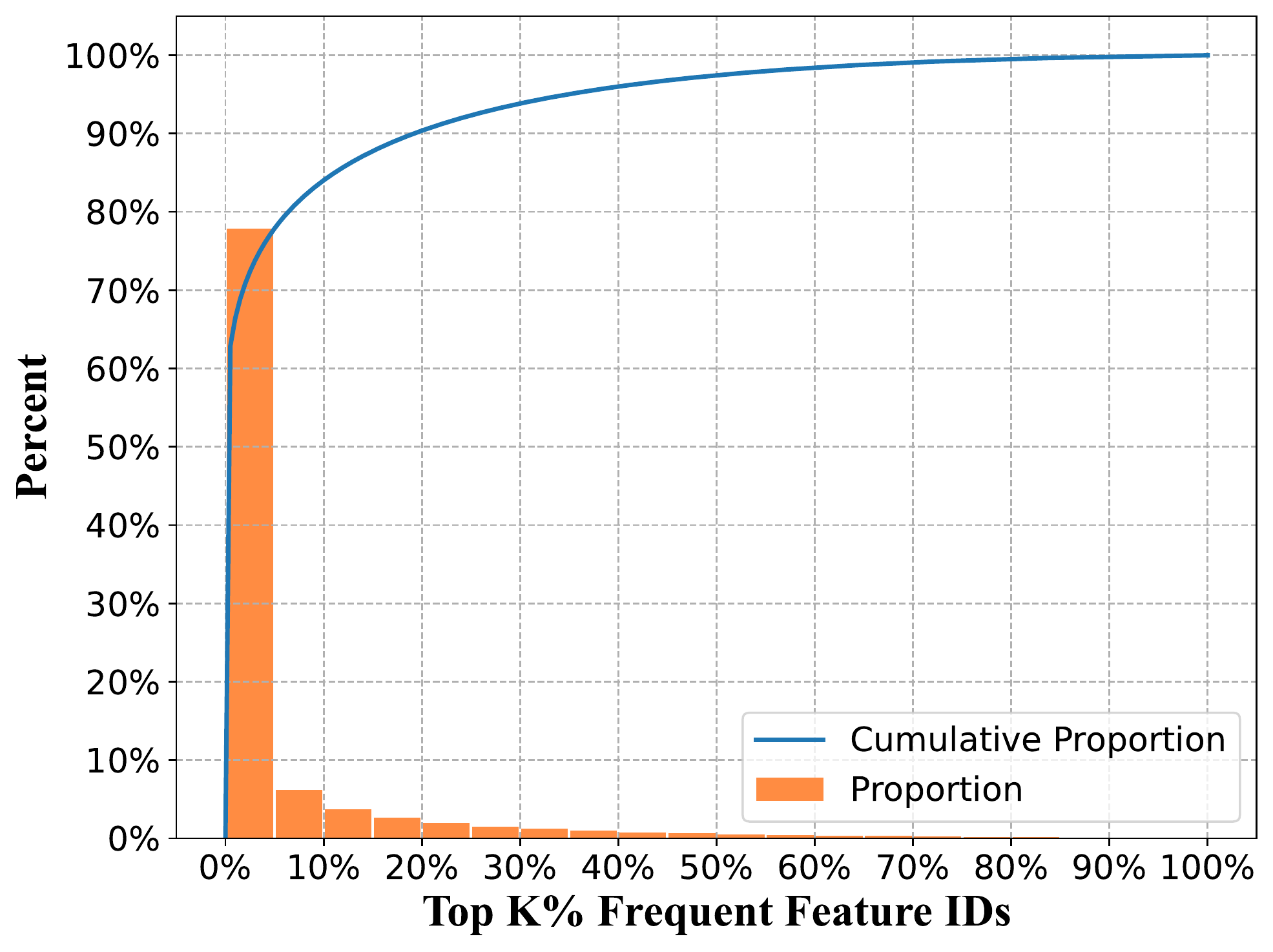}
\centering
\caption{The cumulative proportion of IDs on the production data set. We sort all IDs in descending order by occurrence frequency.}
\label{fig:data_pdf}
\end{figure}
Here, we illustrate the feature sparsity in the production data set. A sparser feature field corresponds to a larger number of unique IDs and a smaller average occurrence per ID. The production data set contains seven feature fields, and Table~\ref{tab:production_feature_field_entropy} gives the number of unique IDs and the mean occurrences of each ID for each feature. The fine-grained features (e.g., item ID and history item IDs) are much sparser than the others. Besides, IDs exhibit a long-tailed distribution. For the whole data set, we sort all feature IDs in descending order according to the occurrence frequency and plot the distribution in Figure~\ref{fig:data_pdf}. We observe that the bottom 50\% frequent IDs only account for the 2.5\% occurrences.

\begin{figure}[t]
\subfigure[Filter]{
\includegraphics[width=.42\columnwidth]{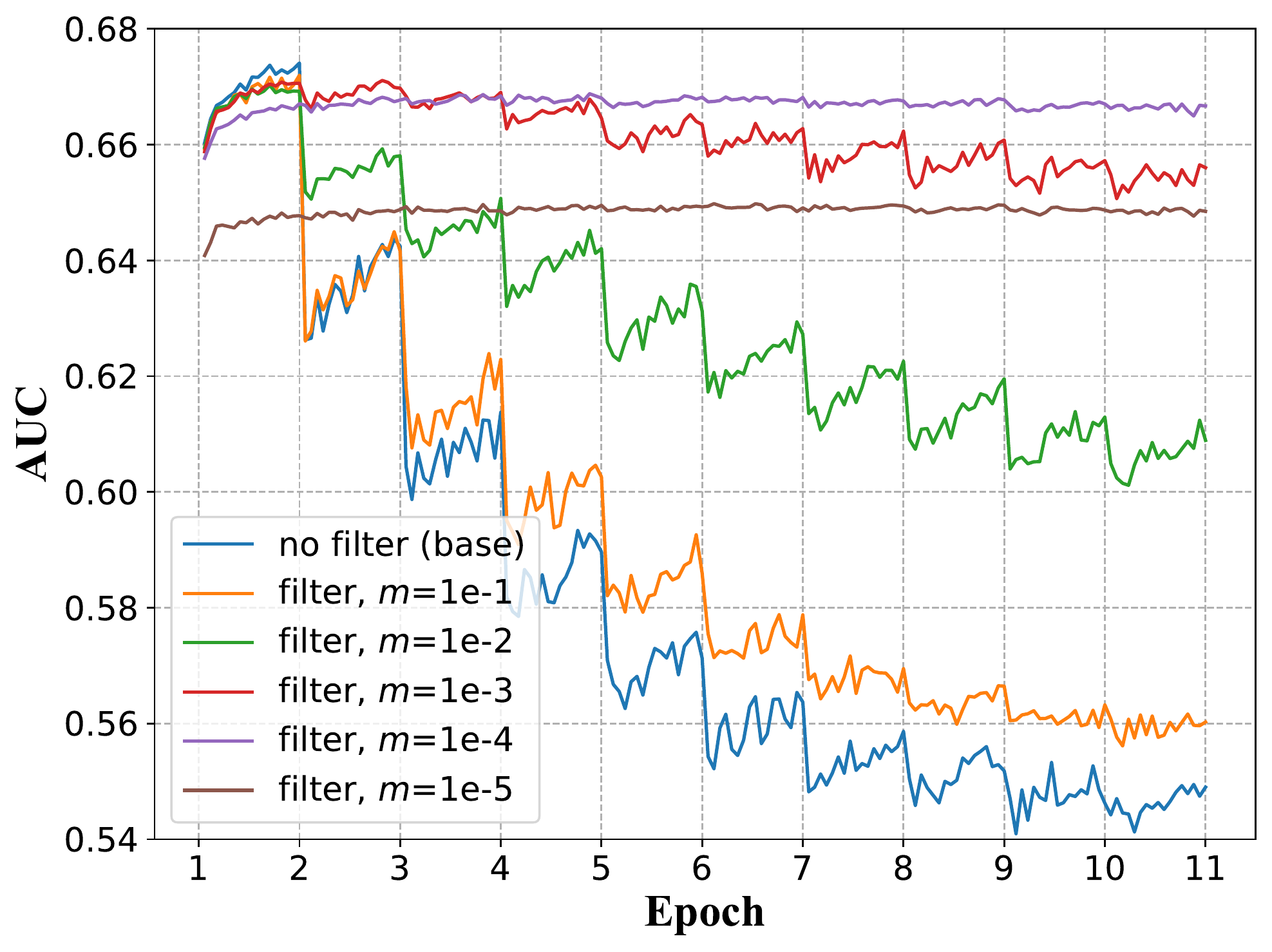}
\label{fig:data_filter_production}
}
\subfigure[Hash]{
\includegraphics[width=.42\columnwidth]{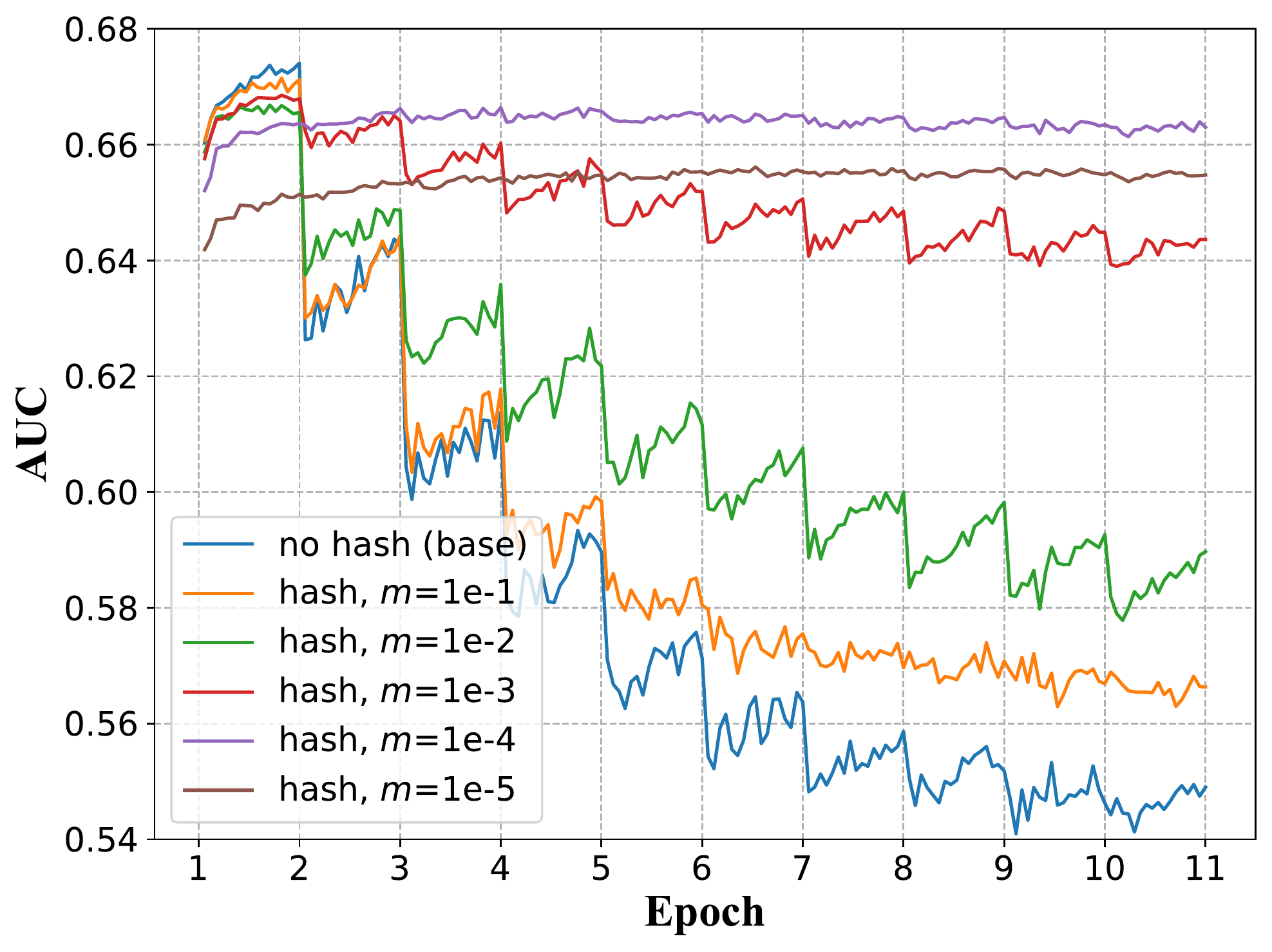}
\label{fig:data_hash_production}
}
\centering
\caption{Test AUCs of models trained on the production data set with filter and hash, respectively. $m$ is the ratio of the number of IDs after and before compression. The smaller $m$, the lower the feature sparsity. The one-epoch phenomenon is related to large feature sparsity.}
\label{fig:data_sparse_production}

\end{figure}

\begin{figure}[t]
\includegraphics[width=.53\columnwidth]{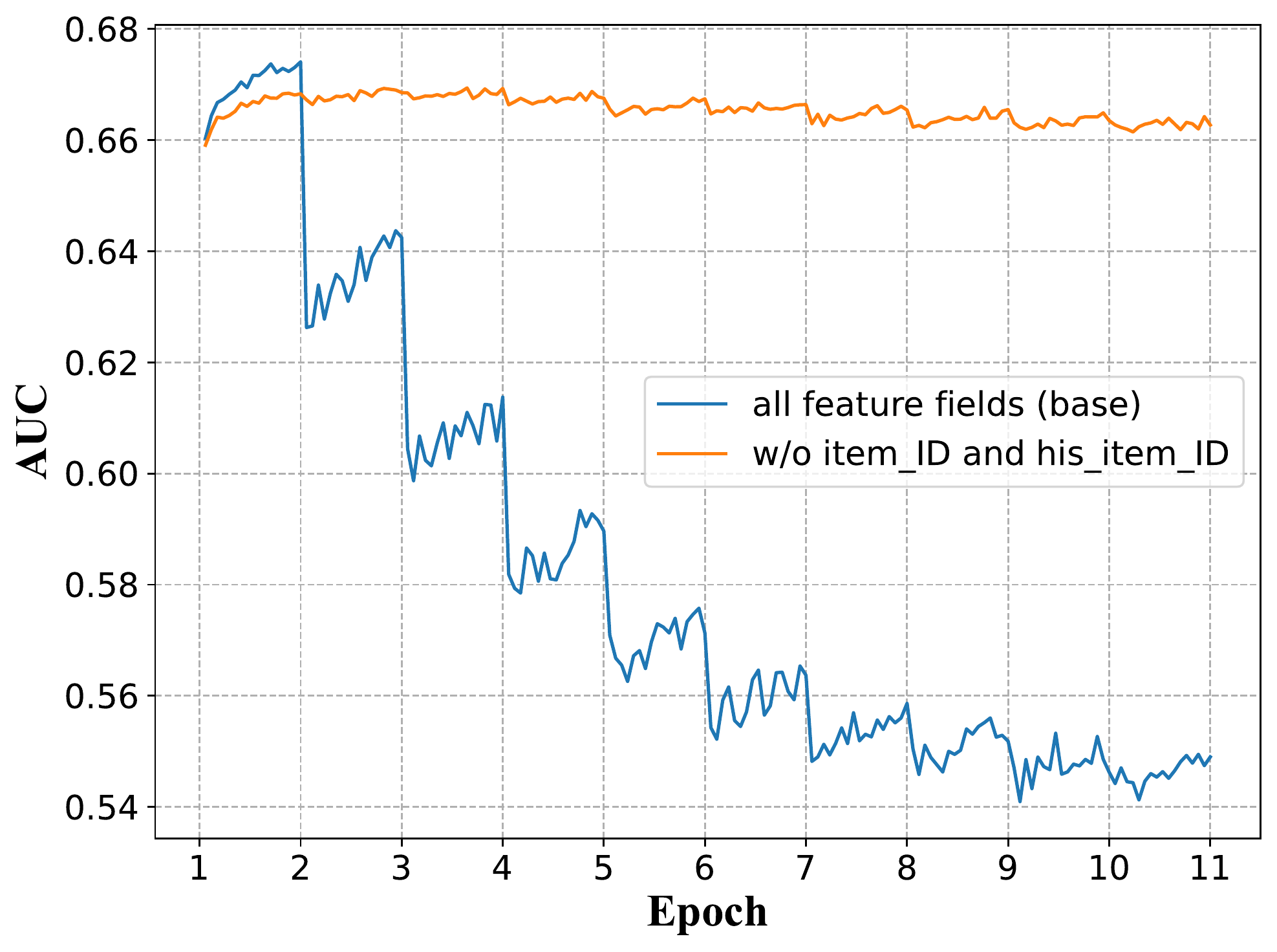}
\centering
\caption{If sparse feature fields (item ID and history item IDs) are not used for training, the model will not experience the one-epoch phenomenon.}
\label{fig:no_sparse_production}

\end{figure}

To show that the feature sparsity is related to the one-epoch phenomenon, we reduce the feature sparsity and illustrate the generalization performance.  We use two techniques, \textbf{filter} and \textbf{hash}, to reduce the feature sparsity. Given a ratio $m$, the filter method only reserves the top $m$ ratio frequent IDs and filters out the other IDs (replacing them with a default ID). As for the hash method, it maps each ID to a space whose size is the $m$ ratio of the number of all IDs. Note that the filter and hash are applied to the whole data set rather than a specific feature field. After filter or hash, the feature sparsity is alleviated. Figure~\ref{fig:data_sparse_production} gives the results, which show that as $m$ gradually reduces, the one-epoch phenomenon is alleviated accordingly. However, at the same time, performance degradation is inevitable.

Another straightforward method to reduce sparsity is to exclude the fine-grained feature fields. In the production data set, we find that when the model does not use item ID and history item IDs, the one-epoch phenomenon does not occur. However, it leads to worse performance than a model that is trained with all features. The result is demonstrated in Figure~\ref{fig:no_sparse_production}.

From the experiment results, it can be concluded that \textbf{the feature sparsity is closely related to the one-epoch phenomenon}. Although the one-epoch phenomenon can be alleviated or even eliminated by reducing feature sparsity, it often leads to inferior model performances. The above experiments also explain why models trained with only coarse-grained features do not encounter the one-epoch phenomenon.

\subsection{Summary and Discussion}
Through extensive experiments, we find that the \textbf{model structure}, \textbf{optimization algorithm with a fast convergence rate} (e.g., Adam optimizer with a large learning rate), and \textbf{large feature sparsity} (e.g., using fine-grained features like item IDs) are closely related to the one-epoch phenomenon. We also verify that some factors have no obvious effect on the one-epoch phenomenon, including the number of model parameters, activation function, batch size, weight decay, and dropout.

\begin{table*}[t]
\centering
\caption{For each feature field of the production data set, below are the number of unique IDs and average number of occurrences of each ID. It's obvious that item\_ID and his\_item\_ID are the two most sparse feature fields. }
\label{tab:production_feature_field_entropy}
\resizebox{.95\textwidth}{!}{%
\centering
\begin{tabular}{c|c c| c c| c c| c| c} 
\toprule
type & \multicolumn{2}{c|}{user feature fields} & \multicolumn{2}{c|}{user behavior sequences} &  \multicolumn{2}{c|}{candidate item feature fields}  & context feature field\\
\hline
feature field & user\_age  & user\_gender  & his\_item\_ID & his\_item\_cate & item\_ID & item\_cate &  scene\_ID &  \textbf{all} \\ 
\hline
unique IDs  & 10         & 3  & 21,925,711  & 17,727 & 880,613 & 8,223 & 19              &\textbf{22,106,604}      \\
\hline

mean occurrences  & 1,007,307 & 3,357,690 & 23 & 28,411  & 11 & 1,224  & 530,161  & \textbf{50}  \\ 

\bottomrule
\end{tabular}
}
\end{table*}

It's worth mentioning that although we can alleviate the one-epoch phenomenon by changing some factors, it also brings a more or less performance degradation. We find the best performance is achieved by training only one epoch. Most industrial recommender systems only train each sample once and our experiments may provide a reasonable explanation for this practice. Given the experiment result, a natural question is whether we can design a new training paradigm that surpasses the model trained with one epoch. As an exploratory, we have tried methods like fine-tuning part of the parameters and learning rate decay in the second epoch. However,  these methods do not exhibit significant performance gain. We hope that these experiments can shed light on future research on training more than one epoch with better performance.

\section{Hypothesis}
\label{sec:hypothesis}
\begin{figure}[t]
\includegraphics[width=.95\columnwidth]{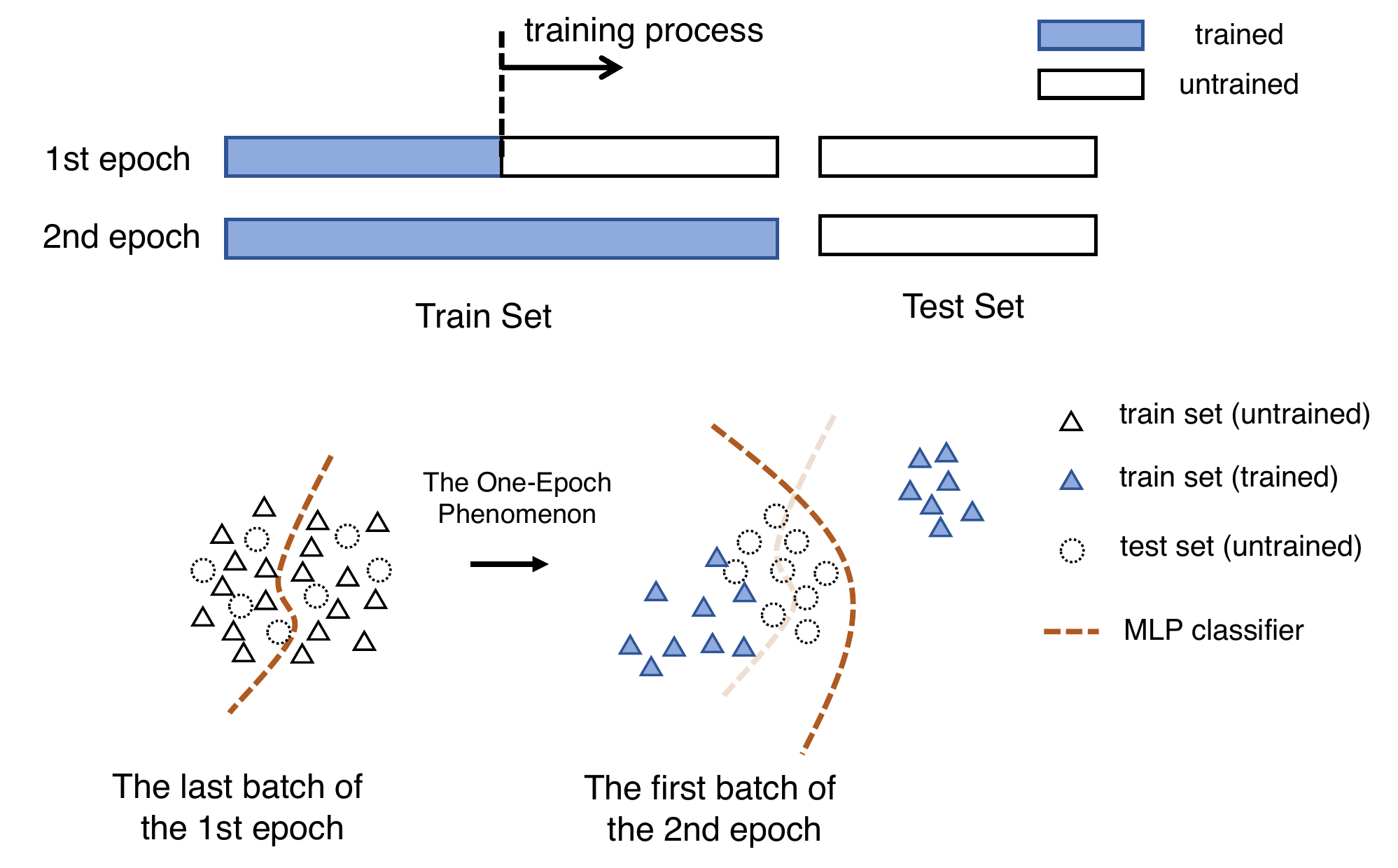}
\centering
\caption{An illustration of the hypothesis. Each $\text{EMB}(\boldsymbol{x})$ is represented by a triangle or a circle. At the beginning of the second epoch, MLP layers quickly fit $\mathcal{D}\!\left(\text{EMB}(\boldsymbol{x_{\text{trained}}}), y\right)$.}
\label{fig:hyp_classifier}
\end{figure}

In this section, we give a hypothesis to explain the one-epoch phenomenon. Let $\text{EMB}(\boldsymbol{x})$ denote the intermediate representation of a sample $\boldsymbol{x}$ after embedding layer. The MLP layers are trained on the joint probability distribution $\mathcal{D}\left(\text{EMB}(\boldsymbol{x}), y\right)$. Denote a trained sample of the model as $\boldsymbol{x}_{\text{trained}}$ and an untrained sample as $\boldsymbol{x}_{\text{untrained}}$. For example, a sample of the train set is $\boldsymbol{x}_{\text{untrained}}$ in the first epoch and $\boldsymbol{x}_{\text{trained}}$ in the second epoch, and a sample of the test set is  $\boldsymbol{x}_{\text{untrained}}$. We hypothesize that $\mathcal{D}\left(\text{EMB}(\boldsymbol{x_{\text{untrained}}}), y\right)$  is significantly different from $\mathcal{D}\left(\text{EMB}(\boldsymbol{x_{\text{trained}}}), y\right)$. At the beginning of the second epoch, MLP layers quickly adapt to the empirical distribution $\mathcal{D}\left(\text{EMB}(\boldsymbol{x_{\text{trained}}}), y\right)$, and the overfitting occurs suddenly, causing the one-epoch phenomenon. Figure~\ref{fig:hyp_classifier} gives an illustration of our hypothesis. In the following, we design a series of proof-of-concept experiments to verify this hypothesis.

\begin{figure}[t]
\centering
\subfigure[All feature fields]{
\includegraphics[width=.42\columnwidth]{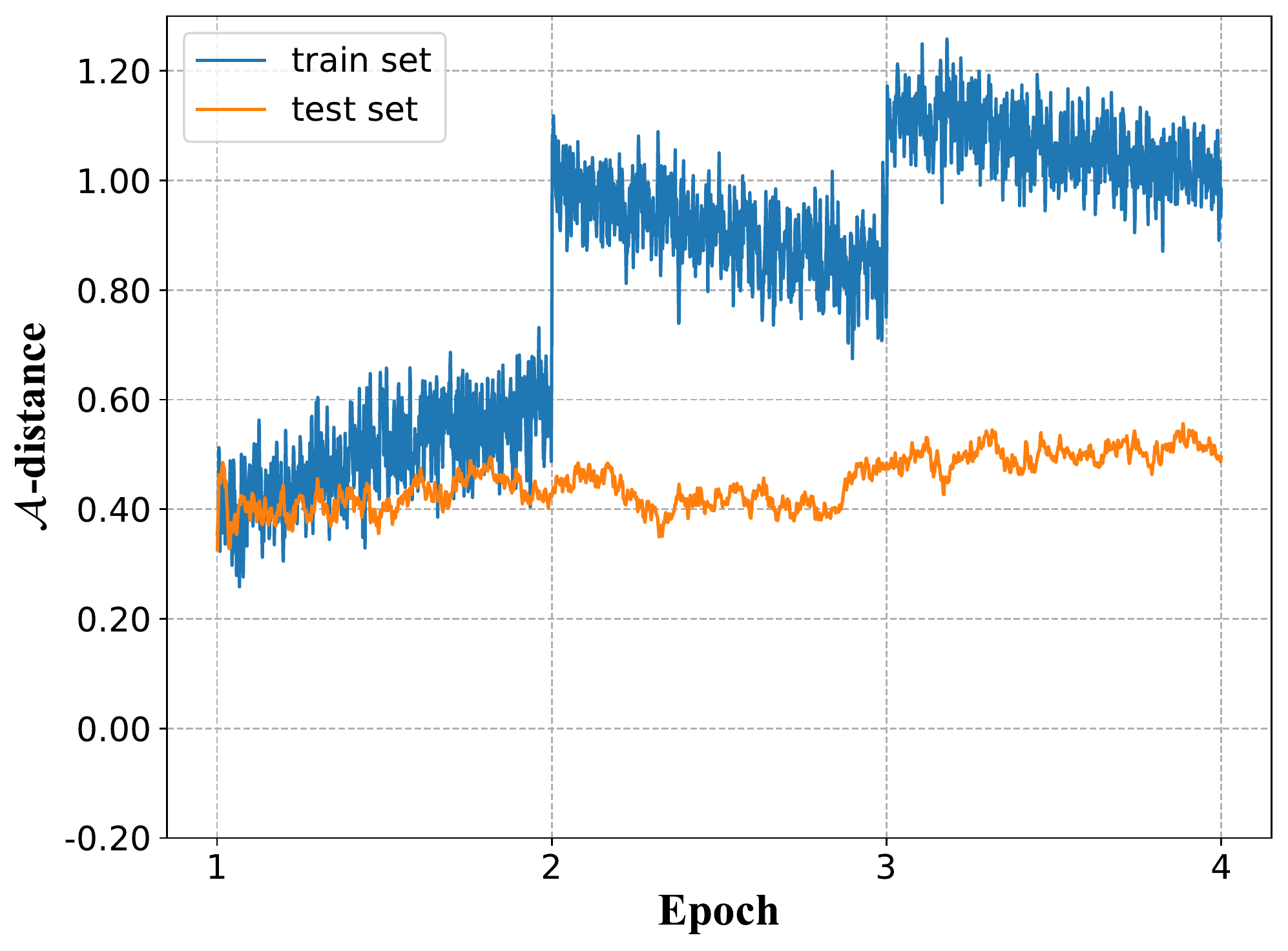}
\label{fig:scores_production_all}

}
\quad
\subfigure[item ID (sparse feature field)]{
\includegraphics[width=.42\columnwidth]{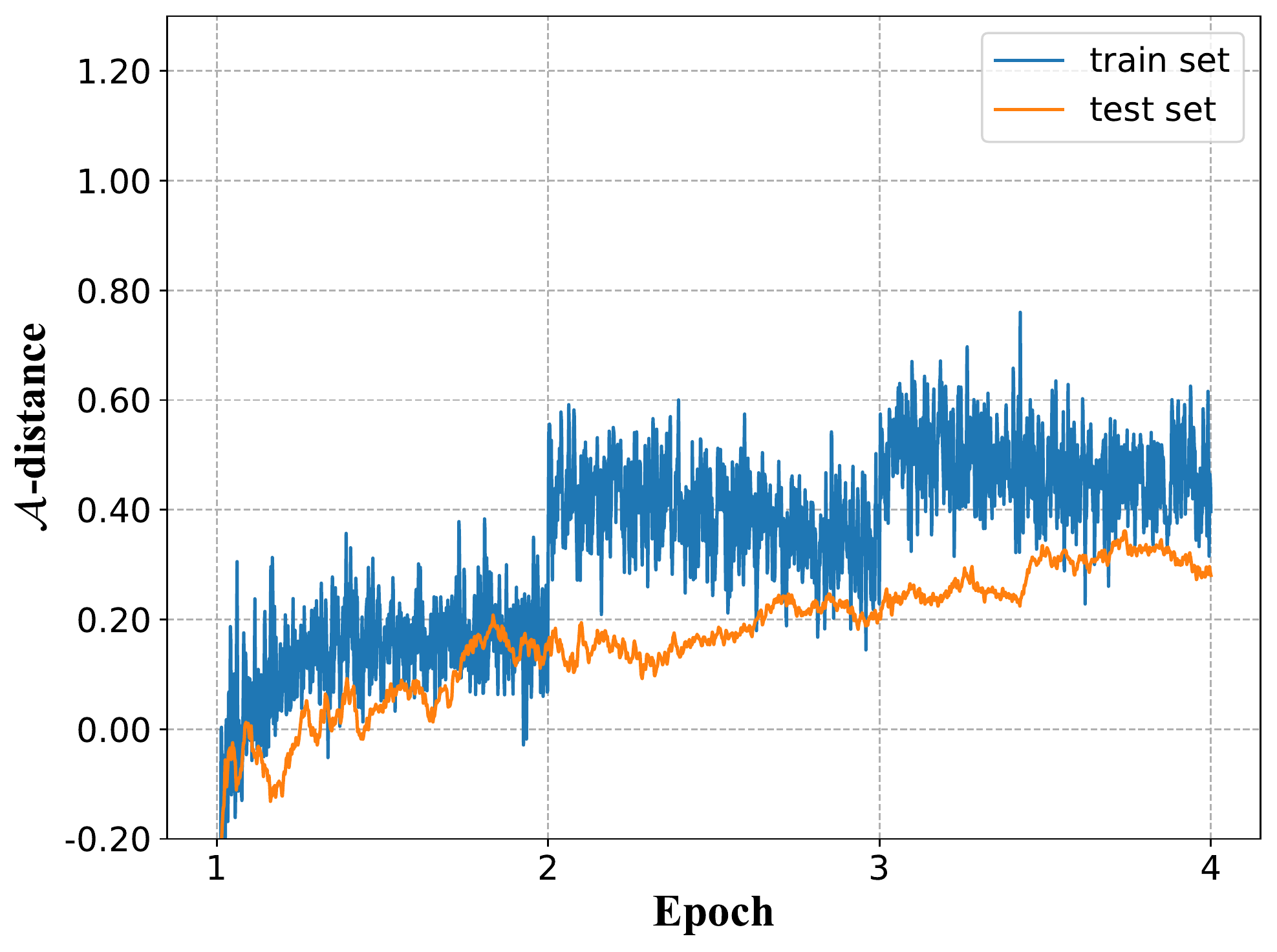}
\label{fig:scores_production_item}

}\\

\subfigure[History item IDs (sparse feature field)]{
\includegraphics[width=.42\columnwidth]{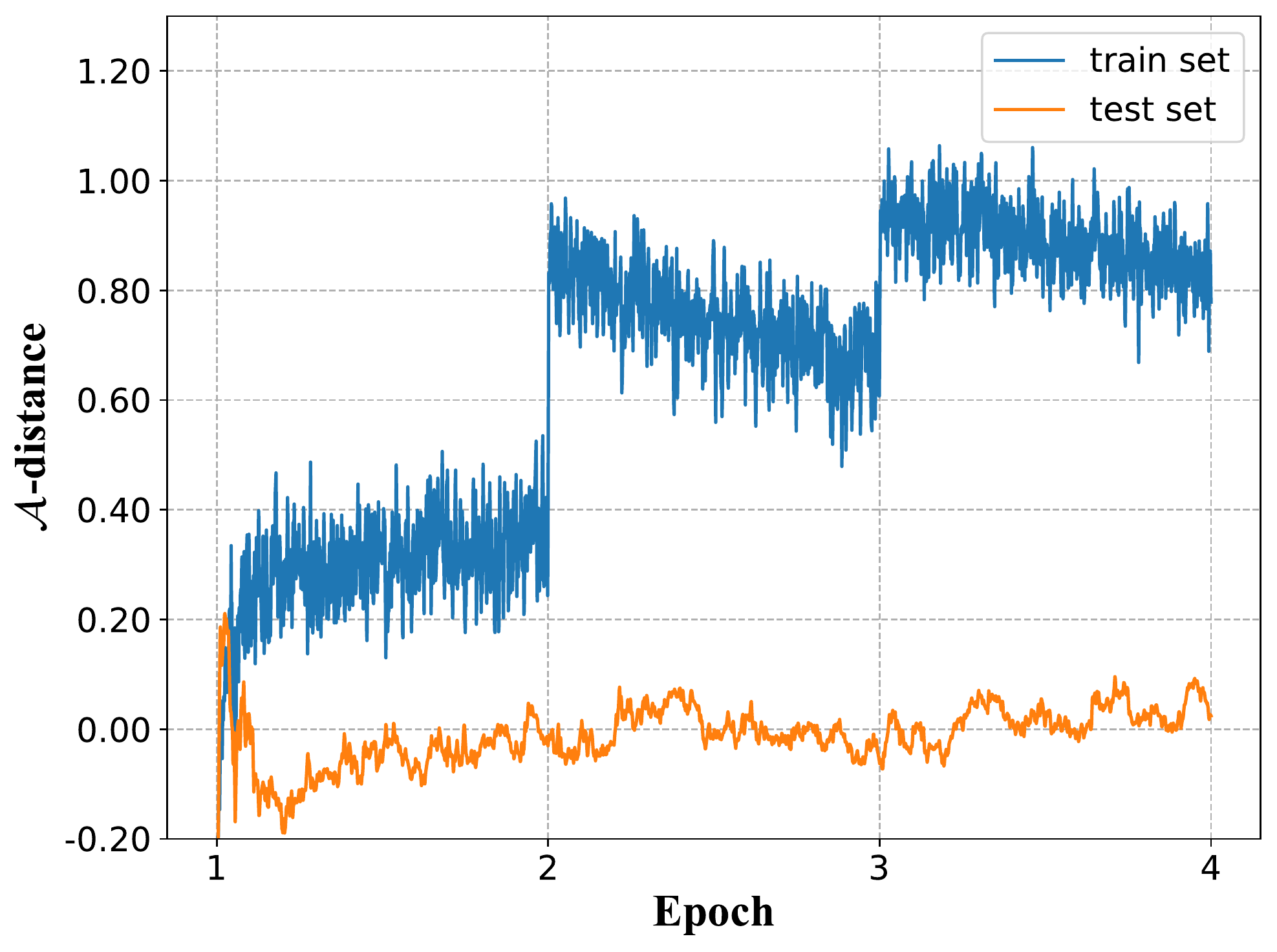}
\label{fig:scores_production_history_item}
}
\quad
\subfigure[History item categories (non-sparse feature field)]{
\includegraphics[width=.42\columnwidth]{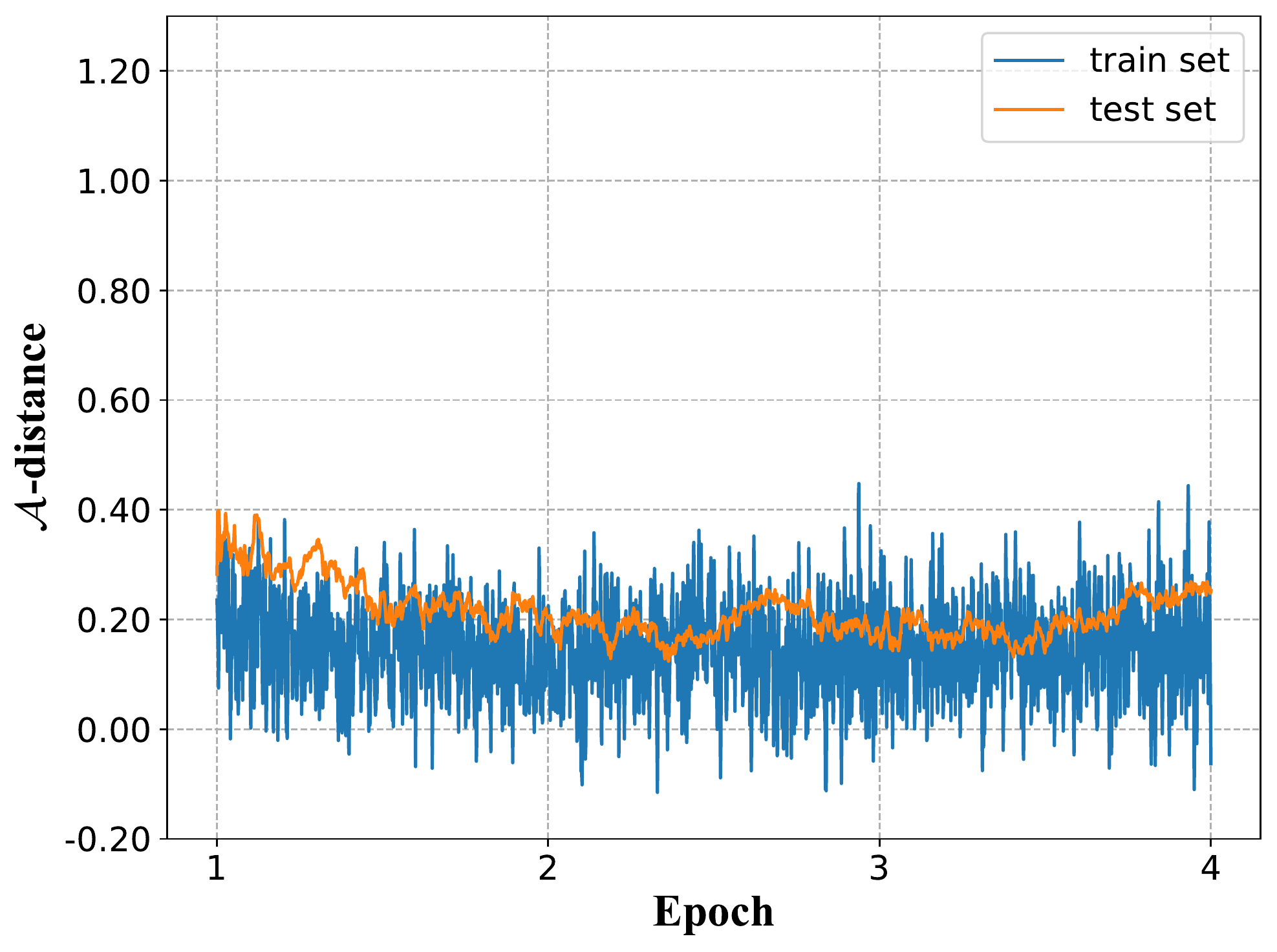}
\label{fig:scores_production_history_cate}
}
\centering
\caption{$\mathcal{A}$-distance between $\mathcal{D}(\text{EMB}(\boldsymbol{x}), y=1)$ and $\mathcal{D}(\text{EMB}(\boldsymbol{x}), y=0)$ in the train set and test set. We use all feature fields or a single feature field to calculate $\mathcal{A}$-distance.}
\label{fig:scores_production}

\end{figure}

\subsection{Difference of the Joint Distribution }
We show that $\mathcal{D}\!\left(\text{EMB}(\boldsymbol{x_{\text{untrained}}}), y\right)$ is significantly different from $ \mathcal{D}\!\left(\text{EMB}(\boldsymbol{x_{\text{trained}}}), y\right)$ . 
However, it is difficult to straightforwardly calculate the change of $\mathcal{D}\left(\text{EMB}(\boldsymbol{x}), y\right)$. To obtain the quantitative variation, we propose to utilize the separability characteristic of the embedding vectors. In the CTR prediction task, the separability characteristic describes the difficulty to distinguish unclicked and clicked samples. Specifically, we represent the separability of unclicked and clicked samples by $\mathcal{A}$-distance\cite{ben2007a_distance} between $\mathcal{D}\left(\text{EMB}(\boldsymbol{x}), y=0\right)$ and $\mathcal{D}\left(\text{EMB}(\boldsymbol{x}), y=1\right)$. To compute the $\mathcal{A}$-distance, we need to train a binary classifier $h$ to distinguish which domain a sample comes from. Let $\text{err}(h)$ represent the loss of the classifier. Denote the $\mathcal{A}$-distance between $\mathcal{D}\left(\text{EMB}(\boldsymbol{x}), y=0\right)$ and $\mathcal{D}\left(\text{EMB}(\boldsymbol{x}), y=1\right)$ as $\mathcal{A}(\mathcal{D}(+,-))$, which is calculated as:
\begin{equation}
\begin{aligned}
\label{eqn:a_distance}
   \mathcal{A}(\mathcal{D}(+,-))&=\mathcal{A}\left(\mathcal{D}\left(\text{EMB}(\boldsymbol{x}), y=0\right), \mathcal{D}\left(\text{EMB}(\boldsymbol{x}), y=1\right)\right)\\
   &=2(1-2 \operatorname{err}(h)).
\end{aligned}
\end{equation}
Note that a larger $\mathcal{A}$-distance means a greater embedding distribution difference between the clicked and unclicked samples, so it is easier to be separated by the MLP classifier.
We can easily measure the change of joint distribution $\mathcal{D}(\text{EMB}(\boldsymbol{x}), y)$ via the change of $\mathcal{A}(\mathcal{D}(+,-))$.

During the training process, we calculate $\mathcal{A}(\mathcal{D}(+,-))$ on train set and test set, respectively. And LR model is used as the binary classifier for $\mathcal{A}$-distance.
The results are shown in Figure~\ref{fig:scores_production_all}. For the train set, a sample is untrained ($\boldsymbol{x}_{\text{untrained}}$) in the first epoch, while it is trained ($\boldsymbol{x}_{\text{trained}}$) in the second epoch. It can be found that $\mathcal{A}(\mathcal{D}(+,-))$ suddenly increases at the beginning of the second epoch, which verifies that $\mathcal{D}\left(\text{EMB}(\boldsymbol{x}_{\text{trained}}),  y\right)$ is different from $\mathcal{D}\left(\text{EMB}(\boldsymbol{x}_{\text{untrained}}),  y\right)$ and $\mathcal{D}\left(\text{EMB}(\boldsymbol{x}_{\text{trained}}),  y\right)$ is much easier to fit. For the test set, all samples are untrained from beginning to the end and the $\mathcal{A}(\mathcal{D}(+,-))$ is stable, which shows that $\mathcal{D}(\text{EMB}(X_\text{untrained}), y)$ has no mutation during the training process.

\begin{figure}[t]
\subfigure[Filter $m=1e\text{-}4$]{
\includegraphics[width=.42\columnwidth]{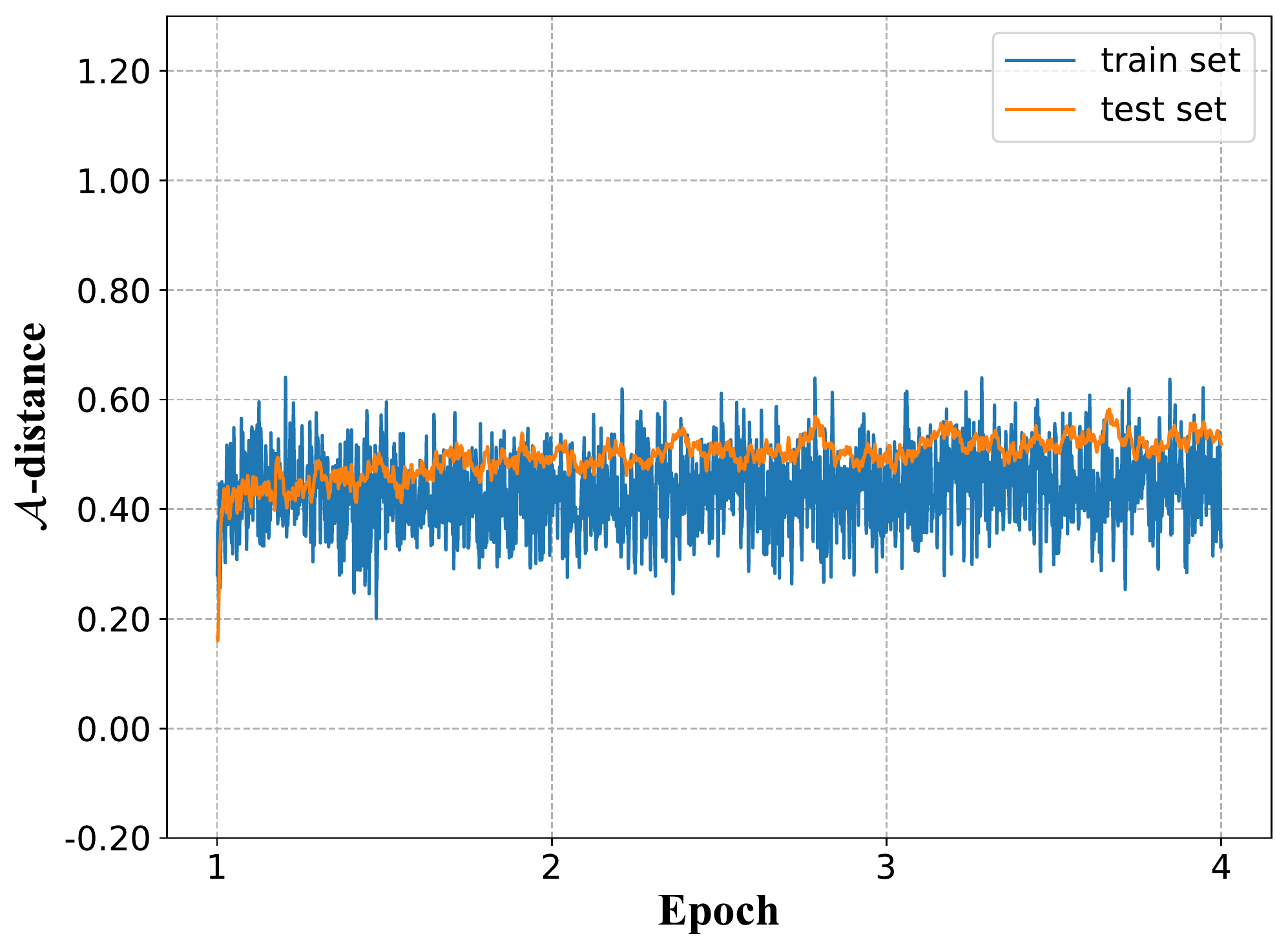}
\label{fig:scores_production_filter_all}
}
\subfigure[Learning rate $1e\text{-}7$]{
\includegraphics[width=.42\columnwidth]{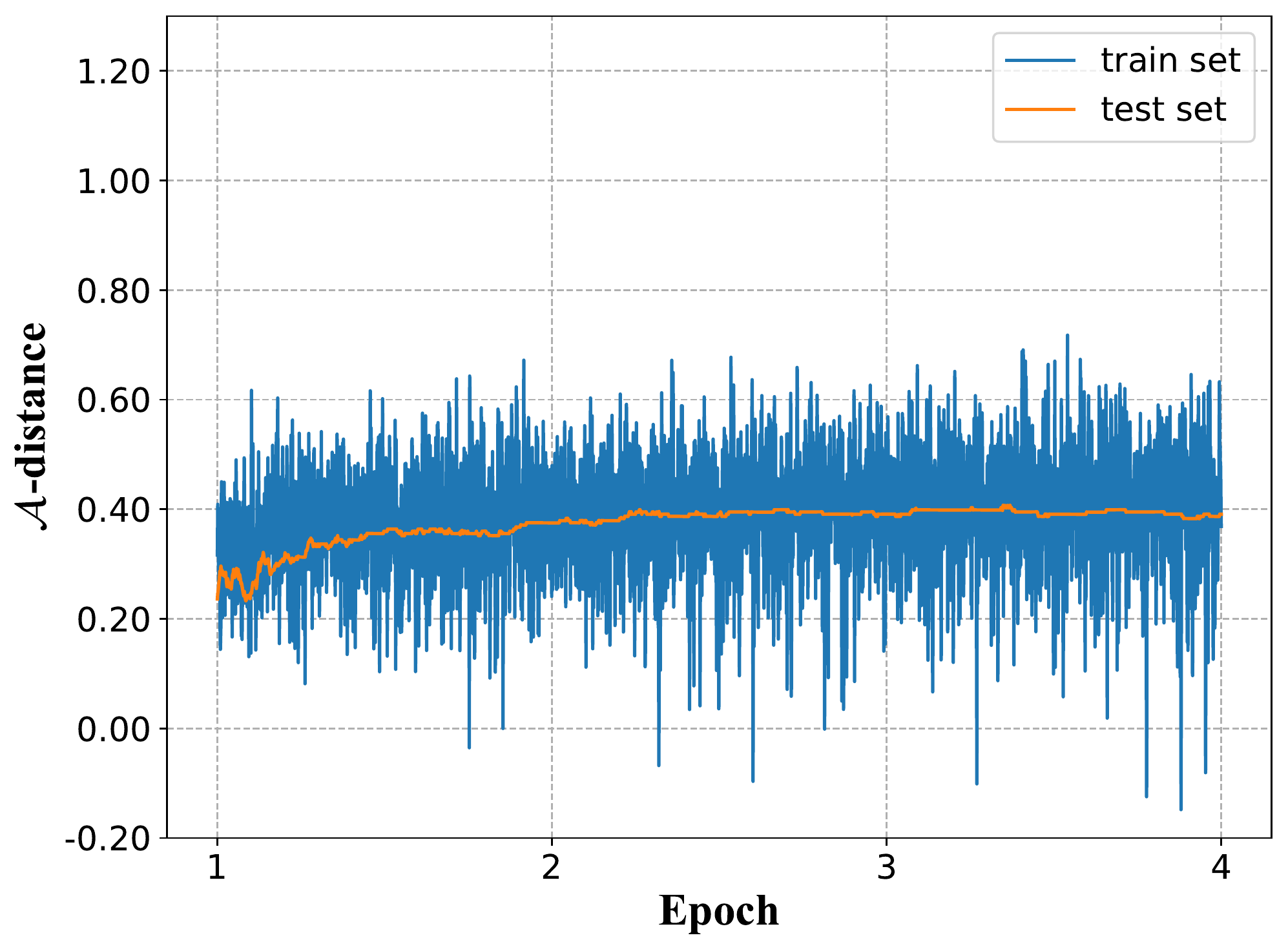}
\label{fig:scores_production_lr1e_7_all}
}
\centering
\caption{When the one-epoch phenomenon does not occur, $\mathcal{A}$-distance between $\mathcal{D}(\text{EMB}(\boldsymbol{x}),y=1)$ and $\mathcal{D}(\text{EMB}(\boldsymbol{x}),y=0)$ in the train set and test set will not suddenly increase.}
\label{fig:scores_production_no_1epoch}

\end{figure}

Further, we try to analyze the relationship between the variation of $\mathcal{D}(\text{EMB}(\boldsymbol{x}), y)$ and the feature sparsity. We follow the above experimental framework but use the embedding vector of each feature field instead of all feature fields to calculate $\mathcal{A}(\mathcal{D}(+,-))$, and other settings of the experiment remain unchanged. We find that the fine-grained feature fields (item ID and history item IDs) have a sudden increase in $\mathcal{A}(\mathcal{D}(+,-))$, while the other features fields (such as age, gender, and history item categories) are not. We show the results of item ID, history item IDs, and history item categories in Figures~\ref{fig:scores_production_item}-\ref{fig:scores_production_history_cate}. This experiment reveals that the difference between $\mathcal{D}(\text{EMB}(\boldsymbol{x}_\text{trained} , y)$ and $\mathcal{D}(\text{EMB}(\boldsymbol{x}_\text{untrained} , y)$ is mainly dominated by the sparse feature fields.

Finally, we also analyze the change of $\mathcal{D}(\text{EMB}(\boldsymbol{x}) , y)$ when the one-epoch phenomenon does not happen. We use learning rate $1e\text{-}7$ and  filter $m=1e\text{-}4$, respectively. Note that the one-epoch phenomenon does not occur (as illustrated in Figure~\ref{fig:model_optimizer_adam} and Figure~\ref{fig:data_filter_production}) in these settings. The result is shown in Figure~\ref{fig:scores_production_no_1epoch}. The curves do not have sudden change between epochs, which indicates that difference between $\mathcal{D}(\text{EMB}(\boldsymbol{x}_\text{trained}), y)$ and $\mathcal{D}(\text{EMB}(\boldsymbol{x}_\text{untrained}), y)$ could be the necessary condition of the one-epoch phenomenon.

\subsection{Fast Adaptation to the Trained Samples}
\begin{figure}[t]
\centering
\subfigure[Embedding layer]{
\includegraphics[width=.42\columnwidth]{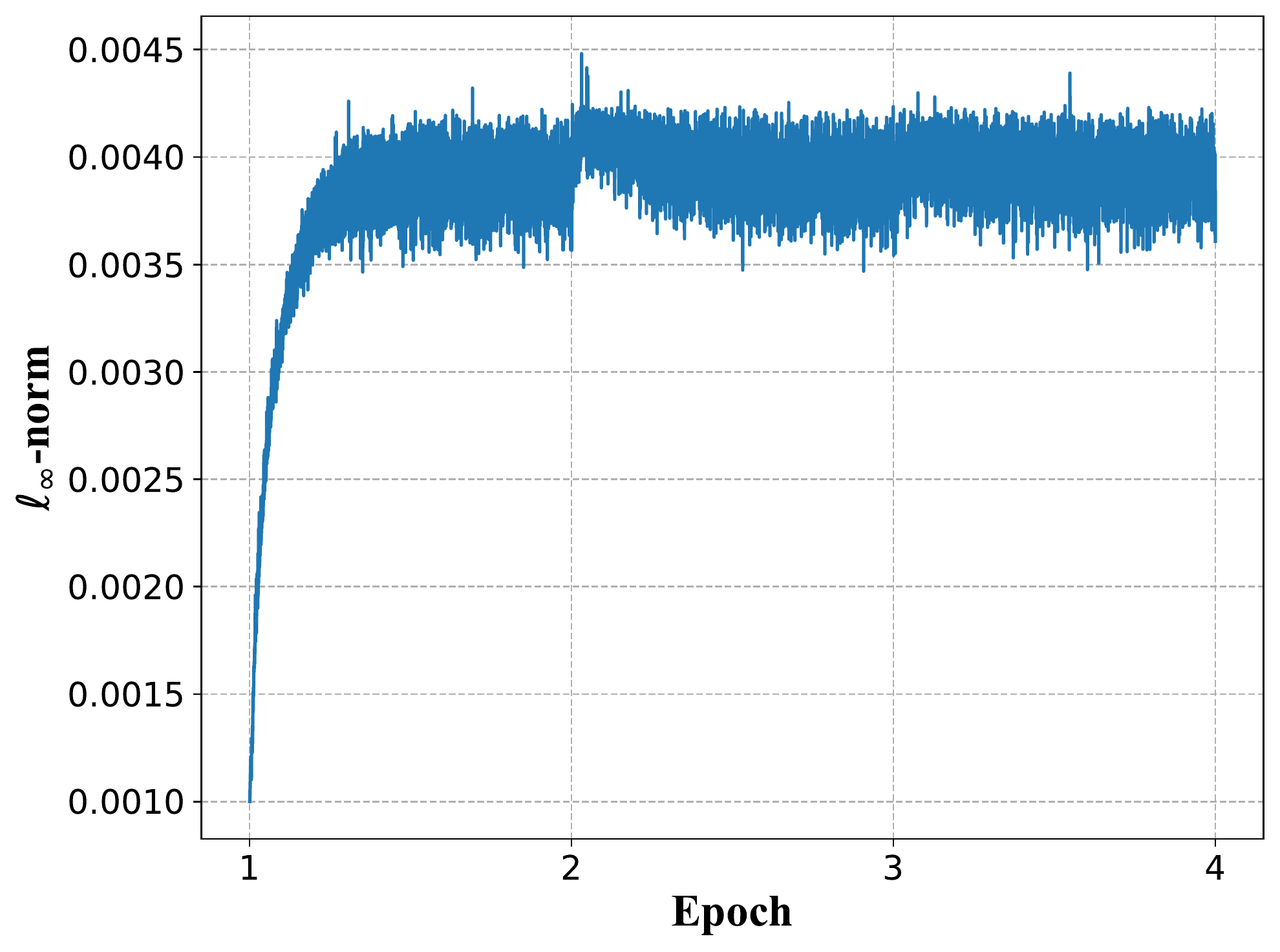}
\label{fig:grad_production_emb}
}
\quad 
\subfigure[FC1]{
\includegraphics[width=.42\columnwidth]{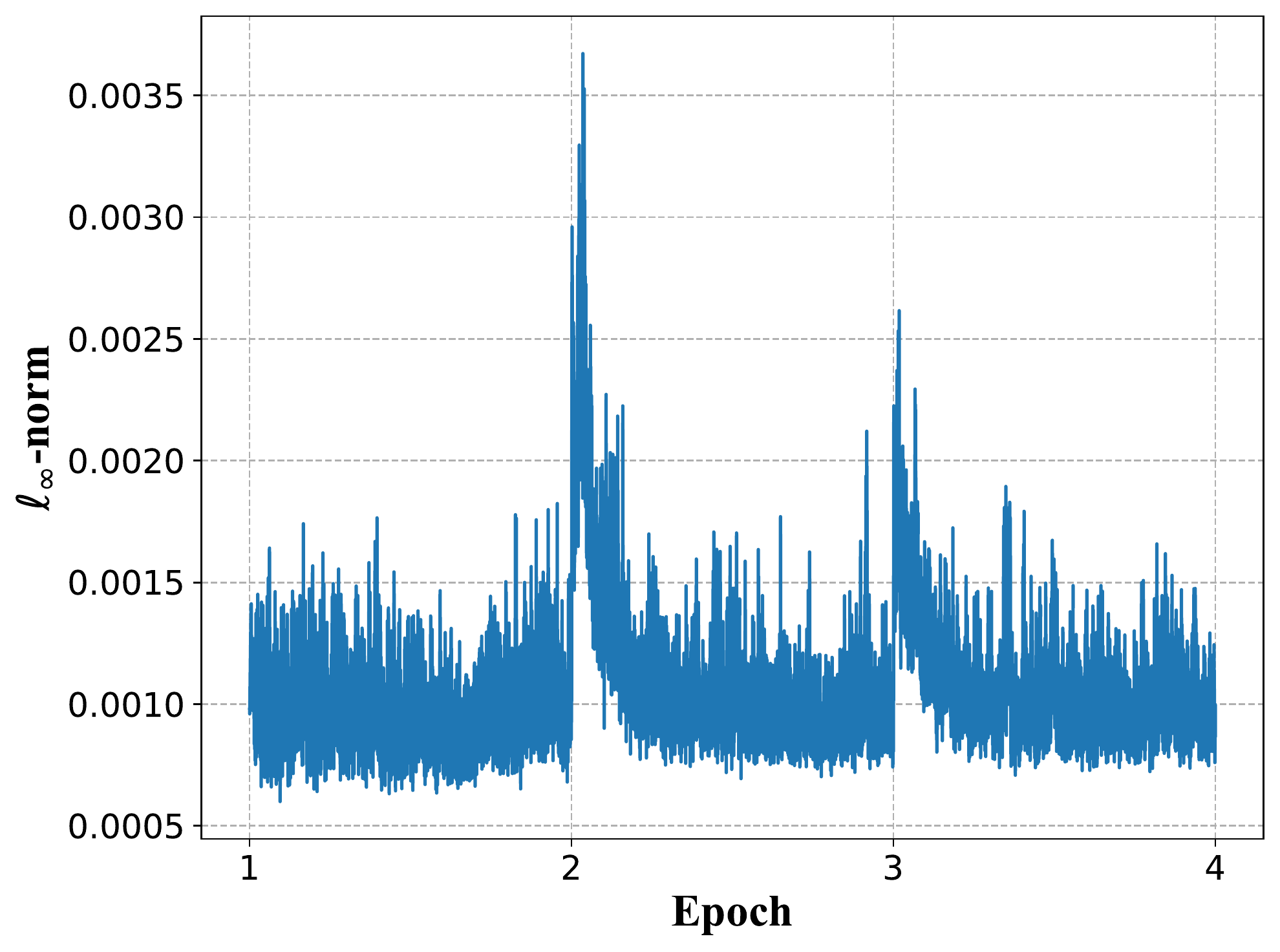}
\label{fig:grad_production_fc1}

}
\\
\subfigure[FC2]{
\includegraphics[width=.42\columnwidth]{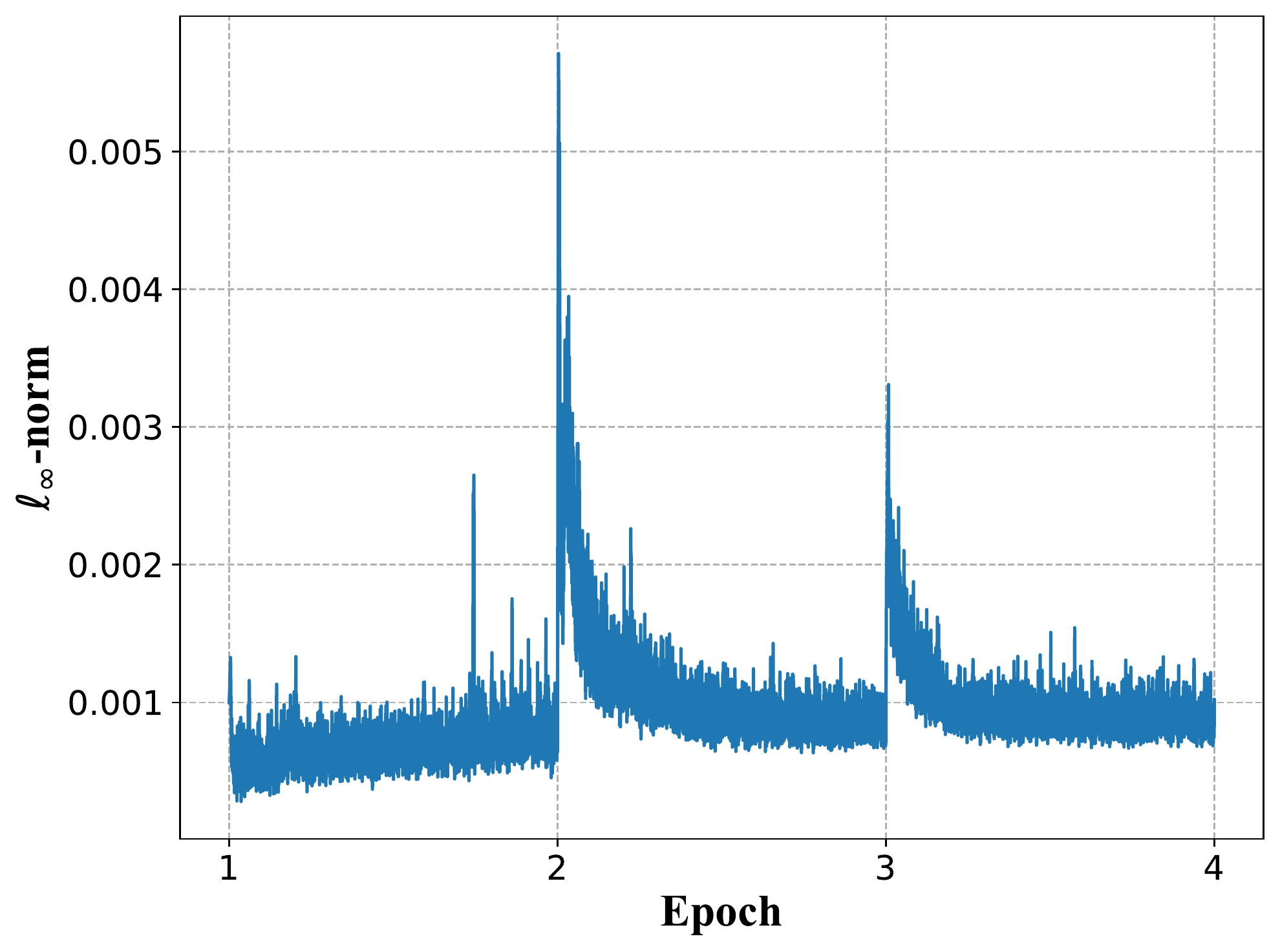}
\label{fig:grad_production_fc2}
}
\quad 
\subfigure[FC3 (output layer)]{
\includegraphics[width=.42\columnwidth]{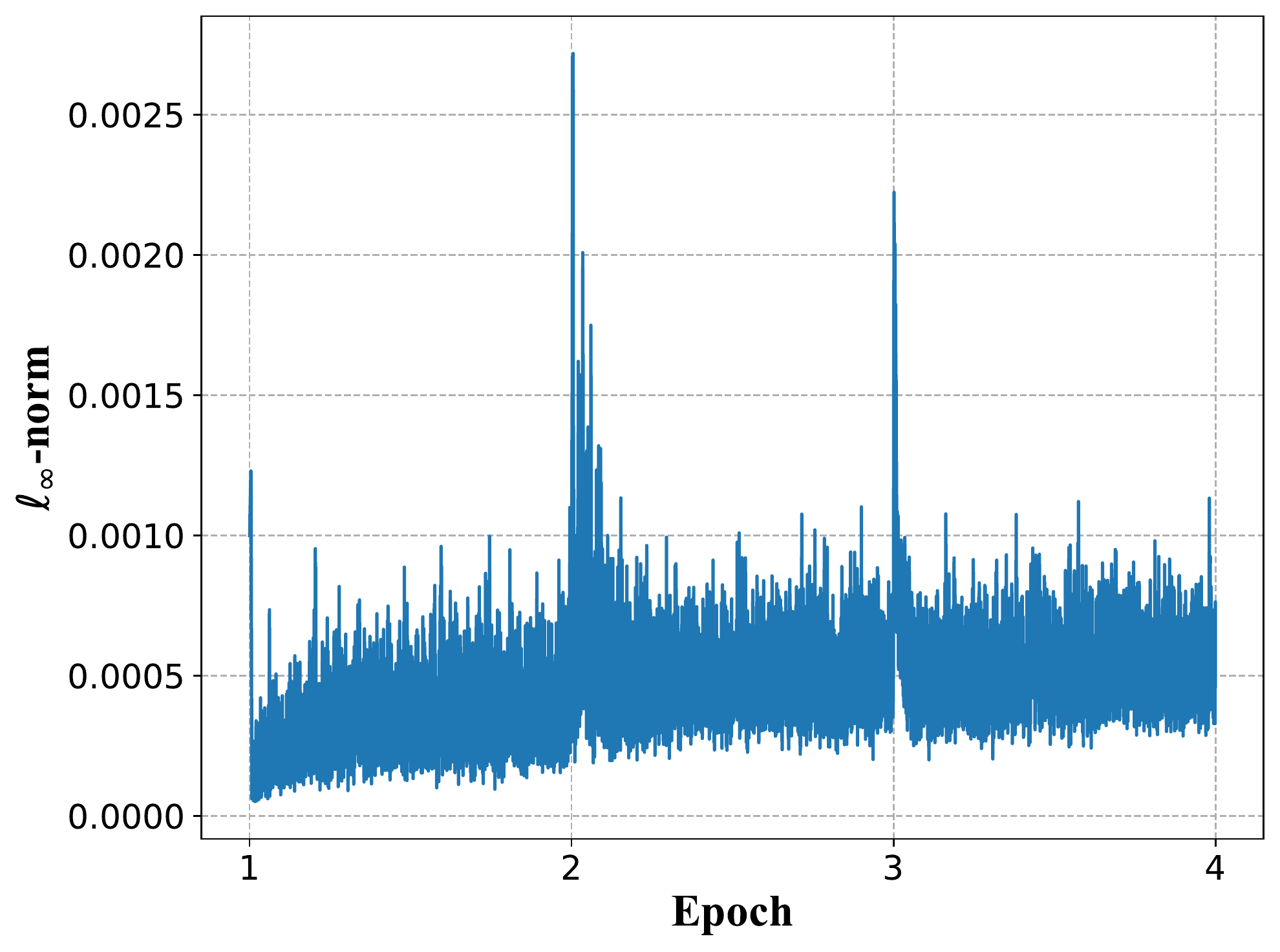}
\label{fig:grad_production_fc3}
}
\centering
\caption{Parameter changes of the Embedding and MLP layers during training. The latter suddenly increases at the beginning of the second epoch, while the former does not. ``FC" is short for ``fully connected".}
\label{fig:grad_production}

\end{figure}

We show that the MLP layers quickly adapt to 
$\mathcal{D}(\text{EMB}(\boldsymbol{x}_\text{trained}), y)$ at the second epoch, in which parameters of the MLP layers change  suddenly.
Particularly, we monitor the parameter changes (i.e., the update values of parameters calculated by the optimizer) of each training step for embedding layer and MLP layers (including the output layer) during the training process. We adopt $\ell_{\infty}$-norm and the results are shown in Figure~\ref{fig:grad_production}. We find that the parameter changes of the embedding layer are generally stable, while the variation of the MLP layers suddenly increases at the second epoch. 

We conduct another experiment to verify the hypothesis that the rapid change of the MLP layers in the second epoch relates to the one-epoch phenomenon. In detail, we fine-tune part of the model parameters after the end of the first epoch,
and freeze the others. The results of fine-tuning all parameters, embedding layers, and MLP layers 
are shown in Figure~\ref{fig:fintune_production}. 
We observe that only fine-tuning the MLP layers leads to the one phenomenon. Only fine-tuning the embedding layer, i.e., freezing the MLP layers after the first epoch, alleviates the one-epoch phenomenon. The result validates the hypothesis that fast adaptation to the trained samples of the MLP layers causes the one-epoch phenomenon.

\subsection{Summary}
According to our hypothesis and the verification experiments, \\ 
$\mathcal{D}\left(\text{EMB}(\boldsymbol{x_{\text{trained}}}), y\right)$ is different from $\mathcal{D}\left(\text{EMB}(\boldsymbol{x_{\text{untrained}}}), y\right)$. The MLP layers quickly adapt to the empirical distribution of trained samples  $\mathcal{D}\left(\text{EMB}(\boldsymbol{x_{\text{trained}}}), y\right)$  at the beginning of the second epoch. Thus, it leads to the two characteristics of the one-epoch phenomenon: (a) it happens exactly at the beginning of the second epoch, and (b) the test performance exhibit a sharp decrease. 

We provide validation experiments to show that the joint distribution $\mathcal{D}\left(\text{EMB}(\boldsymbol{x}), y\right)$ suddenly changes at the beginning of the second epoch via the $\mathcal{A}$-distance metric. And we find the sudden change is mainly caused by fine-grained feature fields. Furthermore, we conduct experiments to show that the parameter update values of MLP layers suddenly increase at the beginning of the second epoch which strongly supports our hypothesis about the MLP layers' fast adaption to the trained samples. And we find that the one-epoch phenomenon can be alleviated via freezing the MLP layers at the second epoch.

\begin{figure}[t]
\includegraphics[width=.53\columnwidth]{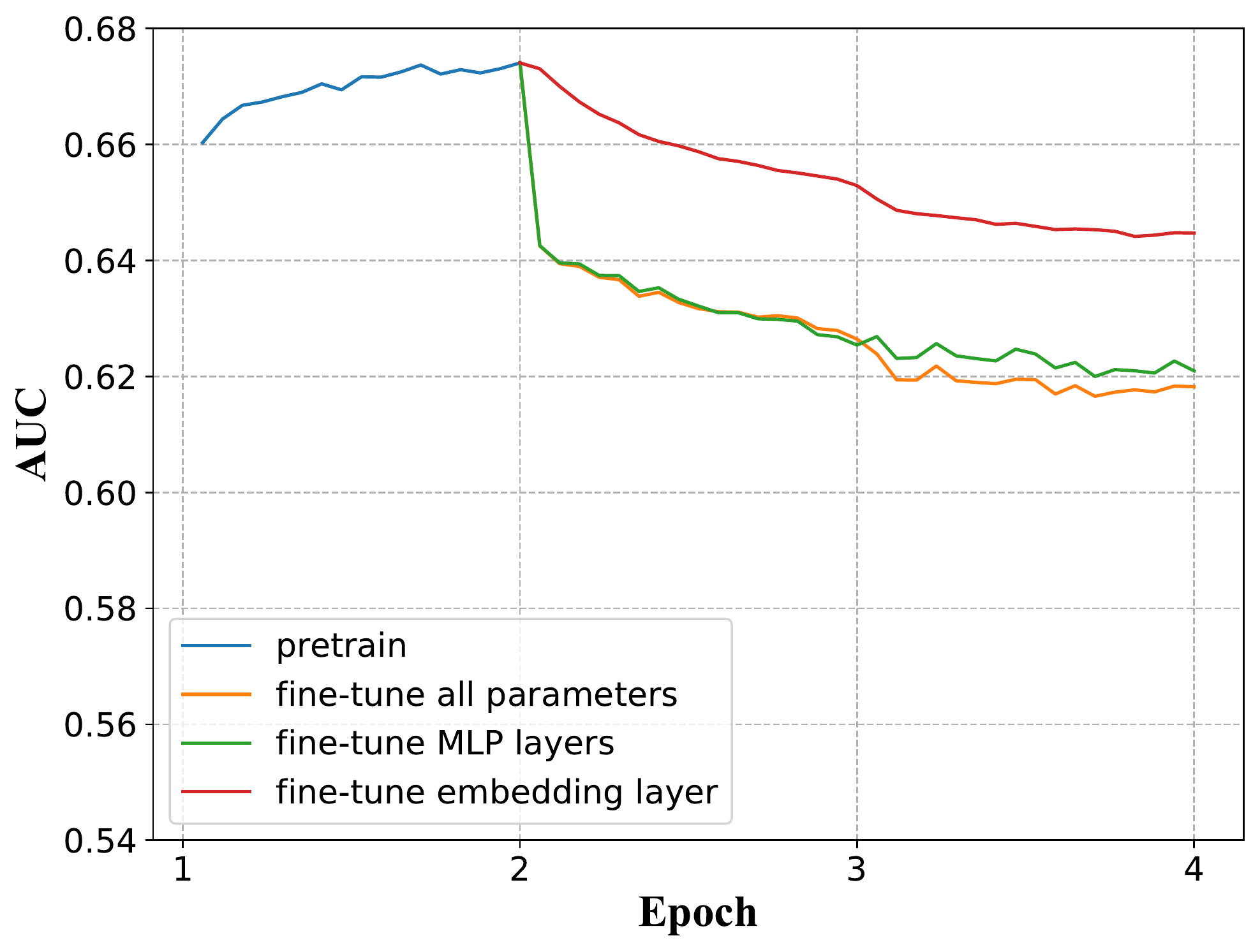}
\centering
\caption{After one epoch of pretraining with default settings, fine-tune part of the model parameters and freeze the others. This figure shows the corresponding test AUCs.}
\label{fig:fintune_production}

\end{figure}

From the experiment results, we can extend our hypothesis: the distribution difference not only exists between the trained and untrained samples but also exists between samples trained at different times. As a result, the test AUC drops rapidly at the beginning of each epoch (from the second epoch), which could be called the ``\textbf{each-epoch phenomenon}". In fact, the experiments in this section (see Figures~\ref{fig:scores_production}, \ref{fig:grad_production}, and \ref{fig:fintune_production}) have verified this more generalized hypothesis. This research mainly focus on the one-epoch phenomenon, and investigate the difference between $\mathcal{D}\left(\text{EMB}(\boldsymbol{x_{\text{trained}}}), y\right)$ and $\mathcal{D}\left(\text{EMB}(\boldsymbol{x_{\text{untrained}}}), y\right)$. We would also like to explore the each-epoch phenomenon in the future.   

\section{Experiment Details}
\label{sec:experiment}

This section describes the details of the data sets and settings. For reproduction, the codes and the results on public data sets are provided at \href{https://github.com/Z-Y-Zhang/one_epoch_phenomenon}{https://github.com/Z-Y-Zhang/one\_epoch\_phenomenon}.

\subsection{Data Sets}
We use one production data set collected from online recommender systems and two widely-used public data sets. 

\textbf{Production} data set is collected from Alibaba display advertisement platform. We randomly select about 1\% of 1-day samples for training and 0.1\% of the next-day samples for testing. 7 feature fields are selected: user age, user gender, history (clicked) item IDs, history (clicked) item categories, item ID, item category, and scene ID. There are 10,073,072 samples and 22,106,604 different IDs in total.  Table~\ref{tab:production_feature_field_entropy} gives the number of unique IDs and the mean occurrences of each ID for each feature field. 


\textbf{Amazon book} data set contains book reviews and metadata from Amazon. Following previous work~\cite{zhou2019dien, BianWRPZXSZCMLX2022CAN}, we regard reviews as positive samples and randomly select products not rated by a specific user as negative
samples, so as to generate the CTR prediction data set. This data set contains 150,016 samples and 425,970 unique IDs in total. The feature fields contain user ID, history (clicked) item IDs, history (clicked) item categories, item ID, and item category. Table~\ref{tab:book_taobao_feature_field_entropy} gives the number of unique IDs and the mean occurrences of each ID for each feature field.

\begin{table*}[t]
\centering
\small
\caption{Number of unique IDs and average number of occurrences of each ID on Amazon book and Taobao data sets. User\_ID, item\_ID and his\_item\_ID are the three most sparse features.}
\label{tab:book_taobao_feature_field_entropy}
\resizebox{.75\textwidth}{!}{%
\begin{tabular}{c|c|cccccc} 
\toprule
\multicolumn{2}{c|}{feature field}         & user\_ID & his\_item\_ID & his\_item\_cate  & item\_ID & item\_cate & \textbf{all}  \\
\hline
\multirow{2}{*}{Amazon book} & unique IDs    & 75,008  & 347,016       & 1,573      & 85,473  & 959     &     \textbf{425,970}     \\
\cline{2-8}
                             & mean occurrences & 2.00   & 19.35         & 4,269.04   & 1.75  & 156.43         & \textbf{32.76}      \\
                             \hline
\multirow{2}{*}{Taobao}      & unique IDs    & $\backslash$  & 4,039,879     & 9,411    & 725,540 & 7,849      & \textbf{4,049,291}        \\
\cline{2-8}
                             & mean occurrences & $\backslash$  & 22.24     & 9,548.70   & 1.36 & 125.8    &    \textbf{44.99}      \\
\bottomrule
\end{tabular}
}
\end{table*}

\textbf{Taobao} data set is a collection of user behaviors (including click, purchase, adding item to shopping cart, and item favoring) from Taobao’s recommender system. Following~\cite{BianWRPZXSZCMLX2022CAN}, we use clicked behaviors for each user to generate a CTR data set. The feature fields contain user ID, history (clicked) item IDs, history (clicked) item categories, item ID, and item category. Because each user has only one sample in this data set, the user ID is useless for training and we exclude this feature field. This data set has 987,648 samples and 4,049,291 unique IDs. The number of unique IDs and the mean occurrences of each ID for each feature field are in Table~\ref{tab:book_taobao_feature_field_entropy}.

\subsection{Settings}
Our base model contains an embedding layer and 3 fully connected layers (200x80x1), that is, 2 MLP layers (200x80) with dice~\cite{zhou2018din} activation and 1 output layer with sigmoid activation. The embedding dimension of each feature field is 8. The batch size is 1024 for the production data set, 128 for Amazon book data set, and 512 for Taobao data set. We use mean pooling to aggregate the user history behavior embedding sequence. The model is optimized by Adam~\cite{KingmaB2014Adam} with learning rate $1e\text{-}3$ to minimize the binary cross-entropy loss. In each experiment, parameters different from the default settings have been described in the context of this paper.


\section{Conclusion and Discussion}
In this paper, we discover that the commonly-adopted deep CTR prediction models exhibit the one-epoch phenomenon: at the beginning of the second epoch, the model performance degrades dramatically, which is a clear sign of overfitting. Such a phenomenon has been witnessed widely in real-world CTR prediction models. Through extensive experiments, we observed that the model structure, optimization algorithm with a fast convergence rate, and feature sparsity are closely related to the one-epoch phenomenon. This explains why online industrial deep CTR prediction models only train the data once. To obtain a better understanding of the one-epoch phenomenon, we propose a likely hypothesis and further validate this premise with a set of experiments. 



Although this research focuses on click-through rate prediction, the above analysis can be easily generalized to other prediction tasks like conversion rate (CVR) prediction. To this end, our findings in this research are of general interest to researchers and practitioners in recommendation systems. We hope that the investigation of the one-epoch phenomenon can shed light on future research on training more epochs for better performance.

\section*{Acknowledgements}
This work is supported by Alibaba Group through Alibaba Innovation Research Program. We would like to thank Guorui Zhou, Xiaoqiang Zhu, Weijie Bian, Zhangming Chan and other colleagues for the helpful discussions.

\bibliographystyle{ACM-Reference-Format}
\bibliography{sample-base}


\end{document}